\ifx\pdfoutput\undefined
	\documentclass[dvips]{ws-procs975x65} 
	\DeclareGraphicsExtensions{.eps}
\else 
	\documentclass[pdftex]{ws-procs975x65} 
	\DeclareGraphicsExtensions{.pdf} 
\fi 

\usepackage{natbib} 

\begin{document} 

\title{Black hole physics and astrophysics: The GRB-Supernova connection and URCA-1 - URCA-2}

\author{R. Ruffini, M.G. Bernardini, C.L. Bianco, L.  Vitagliano and S.-S. Xue}
\address{ICRA -- International Center for Relativistic Astrophysics and Dipartimento di Fisica, Universit\`a di Roma ``La Sapienza'', Piazzale Aldo Moro 5, I-00185 Roma, Italy.}

\author{P. Chardonnet} 
\address{ICRA -- International Center for Relativistic Astrophysics and Universit\'e de Savoie, LAPTH - LAPP, BP 110, F-74941 Annecy-le-Vieux Cedex, France.} 

\author{F. Fraschetti} 
\address{ICRA -- International Center for Relativistic Astrophysics and Universit\`a di Trento, Via Sommarive 14, I-38050 Povo (Trento), Italy.} 

\author{V. Gurzadyan} 
\address{ICRA -- International Center for Relativistic Astrophysics and Yerevan Physics Institute, Alikhanian Brothers Street 2, 375036, Yerevan-36, Armenia} 

\maketitle 

\abstracts{We outline the confluence of three novel theoretical fields in our modeling of Gamma-Ray Bursts (GRBs): 1) the ultrarelativistic regime of a shock front expanding with a Lorentz gamma factor $\sim 300$; 2) the quantum vacuum polarization process leading to an electron-positron plasma originating the shock front; and 3) the general relativistic process of energy extraction from a black hole originating the vacuum polarization process. There are two different classes of GRBs: the long GRBs and the short GRBs. We here address the issue of the long GRBs. The theoretical understanding of the long GRBs has led to the detailed description of their luminosities in fixed energy bands, of their spectral features and made also possible to probe the astrophysical scenario in which they originate. We are specially interested, in this report, to a subclass of long GRBs which appear to be accompanied by a supernova explosion. We are considering two specific examples: GRB980425/SN1998bw and GRB030329/SN2003dh. While these supernovae appear to have a standard energetics of $10^{49}$ ergs, the GRBs are highly variable and can have energetics $10^4$ -- $10^5$ times larger than the ones of the supernovae. Moreover, many long GRBs occurs without the presence of a supernova. It is concluded that in no way a GRB can originate from a supernova. The precise theoretical understanding of the GRB luminosity we present evidence, in both these systems, the existence of an independent component in the X-ray emission, usually interpreted in the current literature as part of the GRB afterglow. This component has been observed by Chandra and XMM to have a strong decay on scale of months. We have named here these two sources respectively URCA-1 and URCA-2, in honor of the work that George Gamow and Mario Shoenberg did in 1939 in this town of Urca identifying the basic mechanism, the Urca processes, leading to the process of gravitational collapse and the formation of a neutron star and a supernova. The further hypothesis is considered to relate this X-ray source to a neutron star, newly born in the Supernova. This hypothesis should be submitted to further theoretical and observational investigation. Some theoretical developments to clarify the astrophysical origin of this new scenario are outlined.}

\section{Introduction} 

In the last century the fundamental discoveries of nuclear physics have led to the understanding of the thermonuclear energy source of main sequence stars and explained the basic physical processes underlying the solar luminosity (see e.g. M. Schwarzschild \cite{msch}).

The discovery of pulsars in 1968 (see Hewish et al. \cite{hbpsc68}) led to the first evidence for the existence of neutron stars, first described in terms of theoretical physics by George Gamow as far back as 1936 \cite{gam36}. It became clear that the pulsed luminosity of pulsars, at times $10^2$ -- $10^3$ larger than solar luminosity, was not related to nuclear burning and could be simply explained in term of the loss of rotational energy of a neutron star (Gold \cite{g68,g69}). For the first time it became so clear the possible relevance of strong gravitational fields in the energetics of an astrophysical system.

\begin{figure} 
\centering 
\includegraphics[width=\hsize,clip]{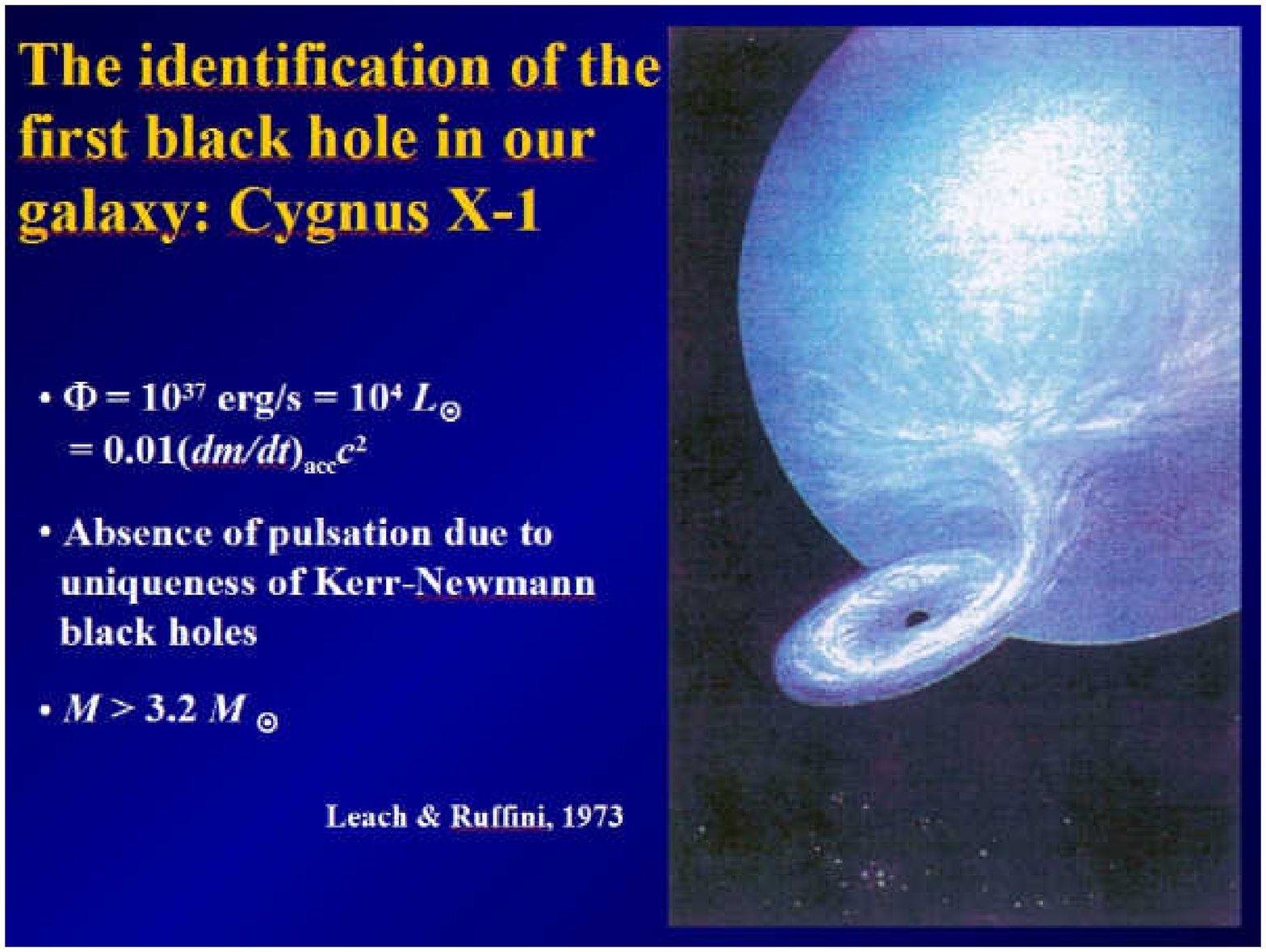} 
\caption{Cygnus X-1 offered the possibility of identifying the first black hole in our galaxy (Leach and Ruffini \cite{lr73}). The luminosity $\Phi$ of $10^4$ solar luminosities points to the accretion process into a neutron star or a black hole as the energy source. The absence of pulsation is naturally explained either by a non-magnetized neutron star or a Kerr-Newmann black hole, which has necessarily to be axially symmetric. What identifies the black hole unambiguously is that the mass of Cygnus X-1, larger than $9M_\odot$, exceeds the absolute upper limit of the neutron star mass, estimated at $3.2M_\odot$ by Rhoades and Ruffini \cite{rr74}.}
\label{CygX1} 
\end{figure}

The birth of X-ray astronomy thanks to Riccardo Giacconi and his group (see e.g. Giacconi and Ruffini \cite{gr78}) led to a still different energy source, originating from the accretion of matter onto a star which has undergone a complete gravitational collapse process: a black hole (see e.g. Ruffini \& Wheeler \cite{rw71}). In this case, the energetics is dominated by the radiation emitted in the accretion process of matter around an already formed black hole. Luminosities up to $10^4$ times the solar luminosity, much larger then the ones of pulsars, could be explained by the release of energy in matter accreting in the deep potential well of a black hole (Leach and Ruffini \cite{lr73}). This allowed to probe for the first time the structure of circular orbits around a black hole computed by Ruffini and Wheeler (see e.g. Landau and Lifshitz \cite{ll2}). This result was well illustrated by the theoretical interpretation of the observations of Cygnus-X1, obtained by the Uhuru satellite and by the optical and radio telescopes on the ground (see Fig. \ref{CygX1}).

The discovery of gamma-ray bursts (GRBs) sign a further decisive progress. The GRBs give the first opportunity to probe and observe a yet different form of energy: the extractable energy of the black hole introduced in 1971 (Christodoulou and Ruffini \cite{cr71}), which we shall refer in the following as the blackholic energy\footnote{This name is the English translation of the Italian words ``energia buconerale'', introduced by Iacopo Ruffini, December 2004, here quoted by his kind permission.}. The blackholic energy, expected to be emitted during the dynamical process of gravitational collapse leading to the formation of the black hole, generates X-ray luminosities $10^{21}$ times larger than the solar luminosity, although lasting for a very short time.

The extreme regimes of GRBs evidence new and unexplored regimes of theoretical physics. It is the aim of this talk to outline the progress achieved in understanding these astrophysical systems and the theoretically predicted regimes for the first time submitted to direct observational verification. 

It is a pleasure to present these results in this village of Urca, in the beautiful city of Rio. While sitting at the Casino de Urca, George Gamow and Mario Schoenberg in 1939 identified the basic process leading to the formation and cooling of a newly born neutron star (see \ref{gamow}). They called this process essentially related to the emission of neutrinos and antineutrinos the Urca process. It is a welcomed coincidence that, in the last hours, while preparing this talk, examining the data of the recently observed GRB 030329, we have received a confirmation of a scenario we have recently outlined in three papers giving the theoretical paradigms for the understanding of GRBs (Ruffini et al. \cite{lett1,lett2,lett3}).

We have clear evidence, first advanced in the system GRB980425/SN1998bw (Ruffini et al. \cite{cospar02}, Fraschetti et al. \cite{f03mg10}) and now confirmed also in the system GRB030329/SN1003dh, that there are in these systems three different components: 1) the GRB source, generated by the collapse to a black hole, 2) the supernova, generated by the collapse of an evolved star, 3) an additional X-ray source which is not related, unlike what is at times stated in the literature, to the GRB afterglow. In honour of the work done in the town of Urca by George Gamow and Mario Schoenberg, identifying in the neutrino emission of the Urca process the basic mechanism leading to the process of gravitational collapse and the formation of a relativistic compact star, we named these two X-ray sources URCA-1, the one formed in the system GRB980425/SN1998bw, and URCA-2, the one formed in the system GRB030329/SN2003dh. We shall now recall some of the main steps in reaching this understanding out of the GRB phenomenon and explore possible explanation of the origin of these two sources.

\section{The energetics of gamma-ray bursts} 

It is well known how GRBs were detected and studied for the first time using the {\em Vela} satellites, developed for military research to monitor the non-violation of the Limited Test Ban Treaty signed in 1963 (see e.g. Strong \cite{s75}). It was clear from the early data of these satellites, which were put at $150,000$ miles from the surface of Earth, that the GRBs did not originate either on the Earth nor in the Solar System.

\begin{figure} 
\centering 
\includegraphics[height=\hsize,clip,angle=90]{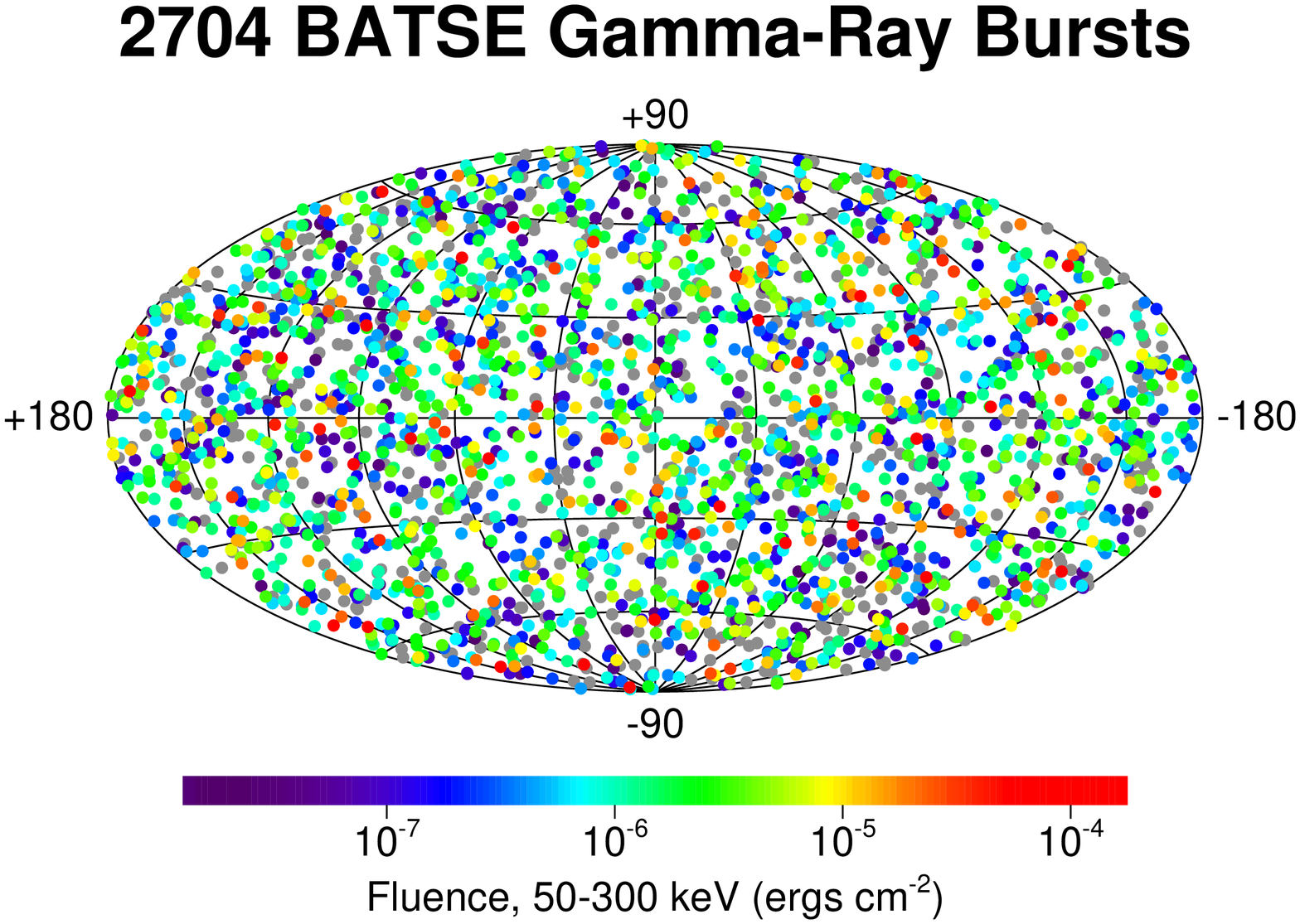} 
\caption{Position in the sky, in galactic coordinates, of 2000 GRB events seen by the CGRO satellite. Their isotropy is evident. Reproduced from BATSE web site by their courtesy.}
\label{batse2k} 
\end{figure} 

\begin{figure} 
\centering 
\includegraphics[width=\hsize,clip]{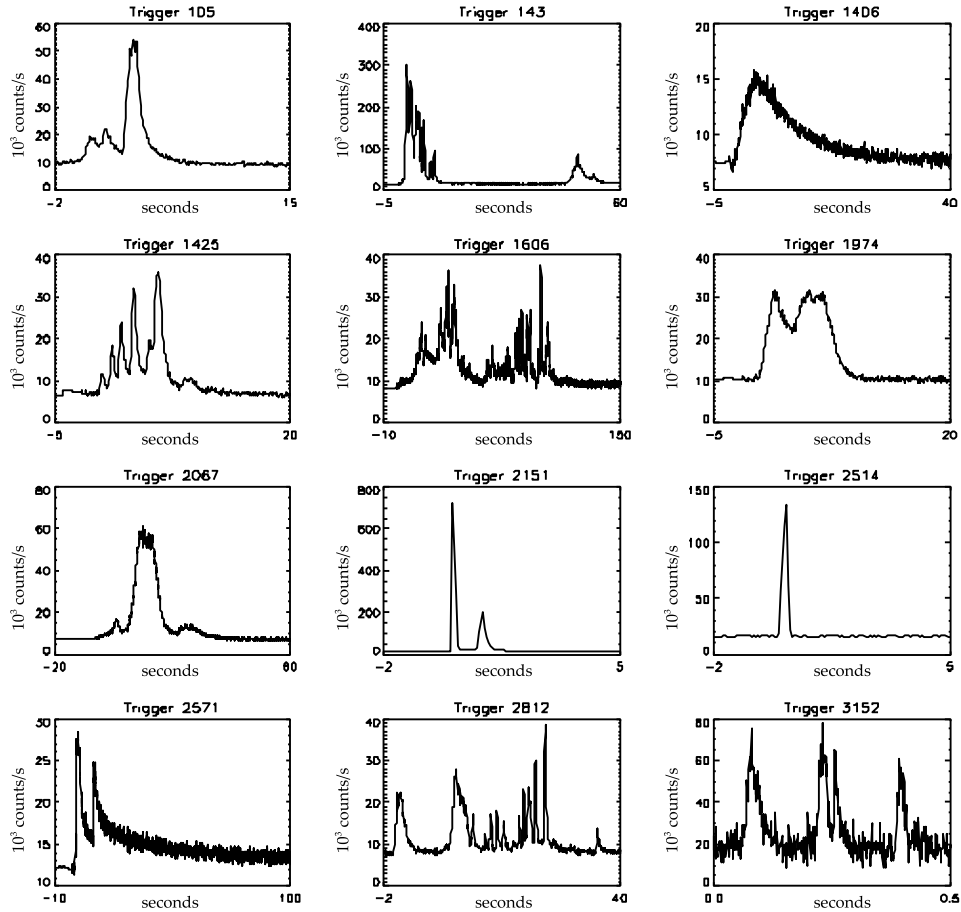} 
\caption{Some GRB light curves observed by the BATSE instrument on board of the CGRO satellite.}
\label{grb_profiles_eng} 
\end{figure}

\begin{figure} 
\centering 
\includegraphics[width=\hsize,clip]{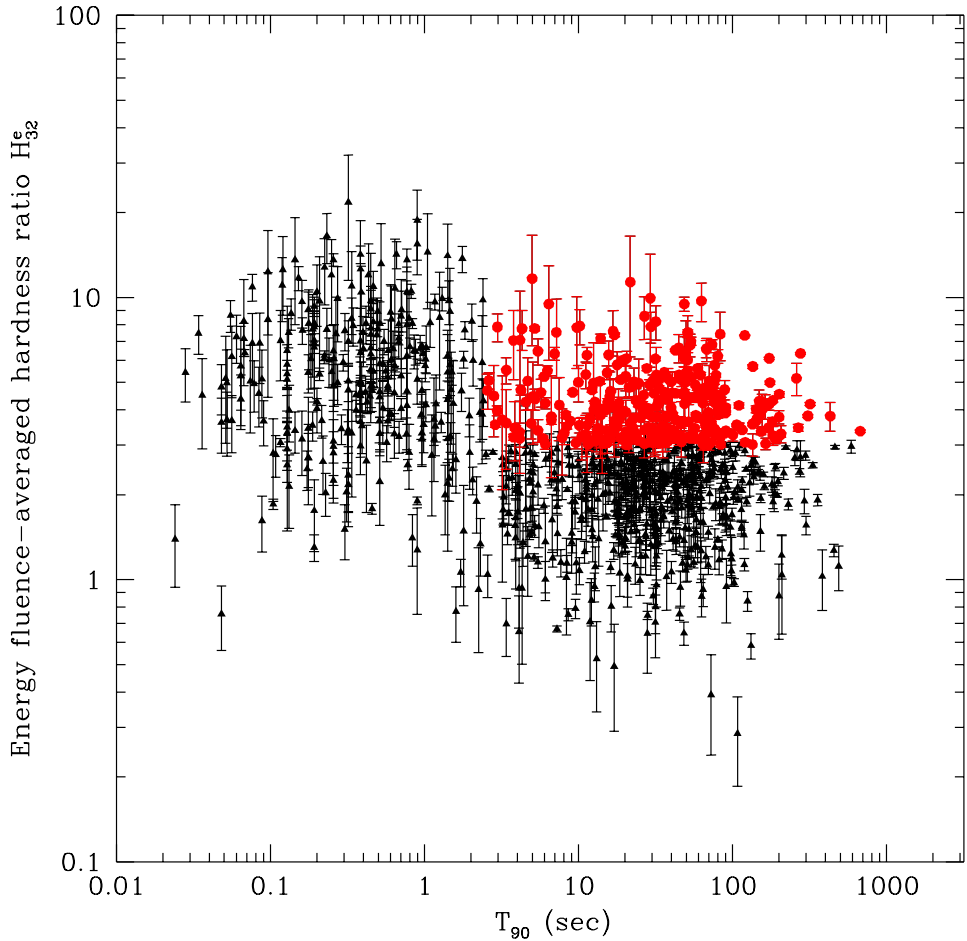} 
\caption{The energy fluence-averaged hardness ratio for short ($T < 1$ s) and long ($T> 1$ s) GRBs are represented. Reproduced, by his kind permission, from Tavani \cite{t98} where the details are given.}
\label{tavani} 
\end{figure}

\begin{figure} 
\centering 
\includegraphics[width=\hsize,clip]{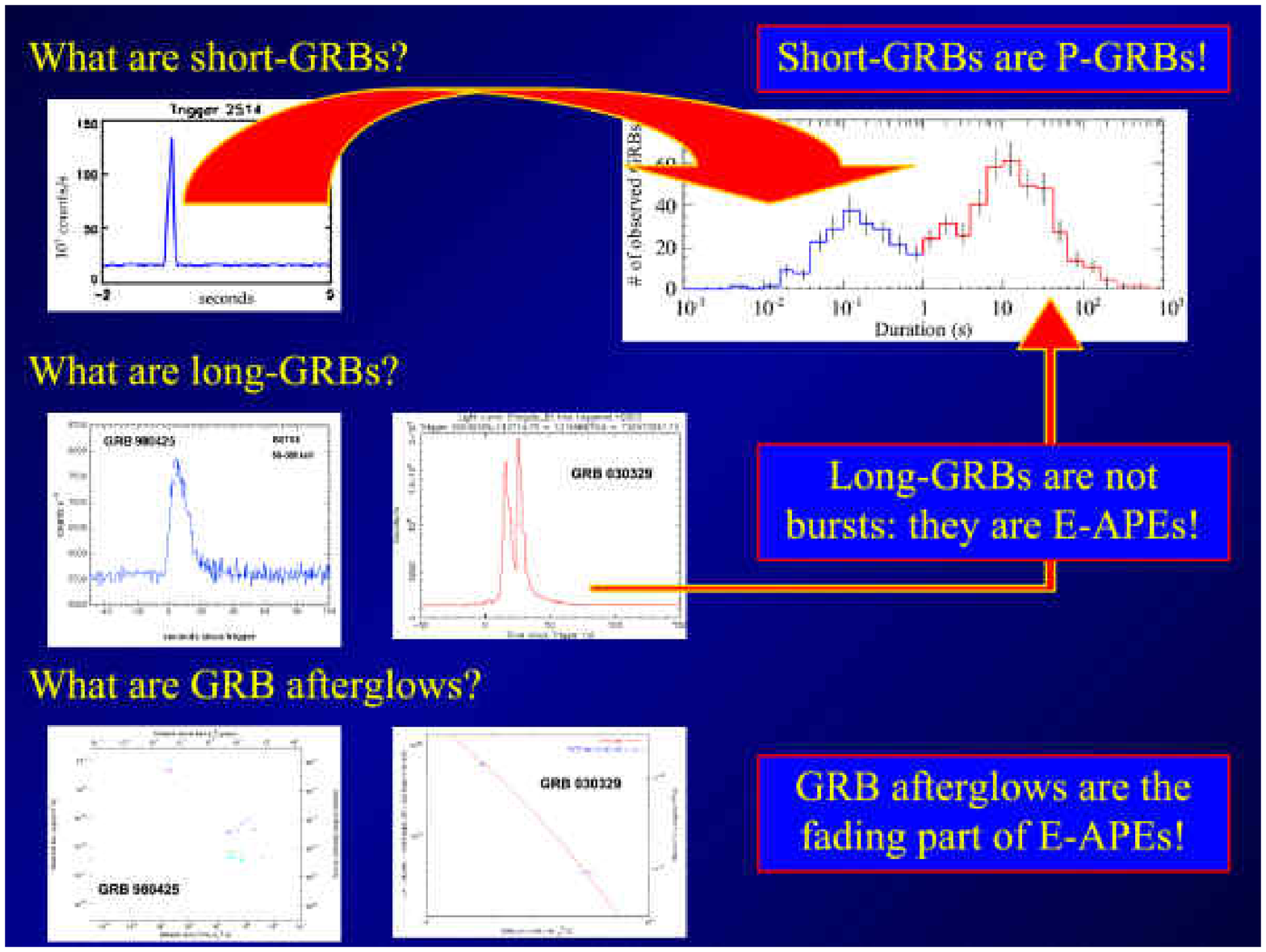} 
\caption{On the upper right part of the figure are plotted the number of the observed GRBs as a function of their duration. The bimodal distribution corresponding respectively to the short bursts, upper left figure, and the long bursts, middle figure, is quite evident. The structure of the long bursts as part of the afterglow phenomena of GRBs is illustrated in section \ref{eape}.}
\label{slb} 
\end{figure}

The mystery of these sources became more profound as the observations of the BATSE instrument on board of the Compton Gamma-Ray Observatory (CGRO) satellite\footnote{see http://cossc.gsfc.nasa.gov/batse/} over $9$ years proved the isotropy of these sources in the sky (See Fig. \ref{batse2k}). In addition to these data, the CGRO satellite gave an unprecedented number of details on the GRB structure, on their spectral properties and time variabilities which became encoded in the fourth BATSE catalog \cite{batse4b} (see e.g. Fig. \ref{grb_profiles_eng}). Out of the analysis of these BATSE sources it soon became clear (see e.g. Kouveliotou et al. \cite{ka93}, Tavani \cite{t98}) the existence of two distinct families of sources: the long bursts, lasting more then one second and softer in spectra, and the short bursts (see Fig. \ref{slb}), harder in spectra (see Fig. \ref{tavani}). We shall return shortly on this topic.

The situation drastically changed with the discovery of the afterglow by the Italian-Dutch satellite BeppoSAX (Costa et al. \cite{ca97}) and the possibility which led to the optical identification of the GRBs by the largest telescopes in the world, including the Hubble Space Telescope, the Keck Telescope in Hawaii and the VLT in Chile, and allowed as well the identification in the radio band of these sources. The outcome of this collaboration between complementary observational technique has led to the possibility of identifying in 1997 the distance of these sources from the Earth and their tremendous energy of the order up to $10^{54}$ erg/second during the burst. It is interesting, as we will show in the following, that an energetics of this magnitude for the GRBs had previously been predicted out of first principles already in 1974 by Damour and Ruffini \cite{dr75}.

The resonance between the X- and gamma ray astronomy from the satellites and the optical and radio astronomy from the ground, had already marked in the seventies the great success and development of the astrophysics of binary X-ray sources (see e.g. Giacconi \& Ruffini \cite{gr78}). This resonance is re-proposed here for GRBs on a much larger scale. The use of much larger satellites, like Chandra and XMM-Newton, and dedicated space missions, like HETE-2 and, in the near future, Swift, and the very fortunate circumstance of the coming of age of the development of unprecedented optical technologies for the telescopes offers opportunities without precedence in the history of mankind. In parallel, the enormous scientific interest on the nature of GRB sources and the exploration, not only of new regimes, but also of totally novel conceptual physical process of the blackholic energy, make the knowledge of GRBs an authentic new frontier in the scientific knowledge.

\section{The complexity an self-consistency of GRB modeling} 

The study of GRBs is very likely ``the'' most extensive computational and theoretical investigation ever done in physics and astrophysics. There are at least three different fields of research which underlie the foundation of the theoretical understanding of GRBs. All three, for different reasons, are very difficult.

The first field of research is the field of special relativity. As I always mention to my students in the course of theoretical physics, this field is paradoxically very difficult since it is extremely simple. In approaching special relativistic phenomena the extremely simple and clear procedures expressed by Einstein in his 1905 classic paper \cite{e05} are often ignored. Einstein makes use in his work of very few physical assumptions, an almost elementary mathematical framework and gives constant attention to a proper operational definition of all observable quantities. Those who work on GRBs use at times very intricate, complex and often wrong theoretical approaches lacking the necessary self-consistency. This is well demonstrated in the current literature on GRBs.

The second field of research essential for understanding the energetics of GRBs deals with quantum electrodynamics and the relativistic process of pair creation in overcritical electromagnetic fields. This topic is also very difficult but for a quite different conceptual reason: the process of pair creation, expressed in the classic works of Heisenberg-Euler-Schwinger \cite{he35,s51} later developed by many others, is based on a very powerful theoretical framework but has never been verified by experimental data. The quest for creating electron-positron pairs by vacuum polarization processes in heavy ion collisions or in lasers has not yet been successfully achieved in Earth-bound experiments (see e.g. Ruffini, Vitagliano, Xue \cite{rvx05}). As we will show here, there is the tantalizing possibility of observing this phenomenon, for the first time, in the astrophysical setting of GRBs on a more grandiose scale.

There is a third field which is essential for the understanding of the GRB phenomenon: general relativity. In this case, contrary to the case of special relativity, the field is indeed very difficult, since it is very difficult both from a conceptual, technical and mathematical point of view. The physical assumptions are indeed complex. The entire concept of geometrization of physics needs a  new conceptual approach to the field. The mathematical complexity of the pseudo-Riemannian geometry contrasts now with the simple structure of the pseudo-Euclidean Minkowski space. The operational definition of the observable quantities has to take into account the intrinsic geometrical properties and also the cosmological settings of the source. With GRBs we have the possibility to follow, from a safe position in an asymptotically flat space at large distance, the formation of the horizon of a black hole with all the associated relativistic phenomena of light bending and time dilatation. Most important, as we will show in details in this presentation, general relativity in connection with quantum phenomena offers, with the blackholic energy, the explanation of the tremendous GRB energy sources.

For these reasons GRBs offer an authentic new frontier in the field of physics and astrophysics. It is appropriate to mention some of the goals of such a new frontier in the above three fields. We recall in the special relativity field, for the first time, we observe phenomena occurring at Lorentz gamma factors of approximately $300$. In the field of relativistic quantum electro-dynamics we see for the first time the interchange between classical fields and the created quantum matter-antimatter pairs. In  the field of general relativity also for the first time we can test the blackholic energy which is the basic energetic physical variable underlying the entire GRB phenomenon.

The most appealing aspect of this work is that, if indeed these three different fields are treated and approached with the necessary technical and scientific maturity, the model which results has a very large redundancy built-in. The approach requires an unprecedented level of self-consistency. Any departures from the correct theoretical treatment in this very complex system lead to exponential departures from the correct solution and from the correct fit of the observations. 

It is so that, as the model is being properly developed and verified, its solution will have existence and uniqueness.

\subsection{GRBs and special relativity} 

The ongoing dialogue between our work and the one of the workers on GRBs, rests still on some elementary considerations presented by Einstein in his classic article of 1905 \cite{e05}. These considerations are quite general and even precede Einstein's derivation, out of first principles, of the Lorentz transformations. We recall here Einstein's words: ``We might, of course, content ourselves with time values determined by an observer stationed together with the watch at the origin of the co-ordinates, and co-ordinating the corresponding positions of the hands with light signals, given out by every event to be timed, and reaching him through empty space. But this co-ordination has the disadvantage that it is not independent of the standpoint of the observer with the watch or clock, as we know from experience''. 

\begin{figure} 
\centering 
\includegraphics[width=\hsize,clip]{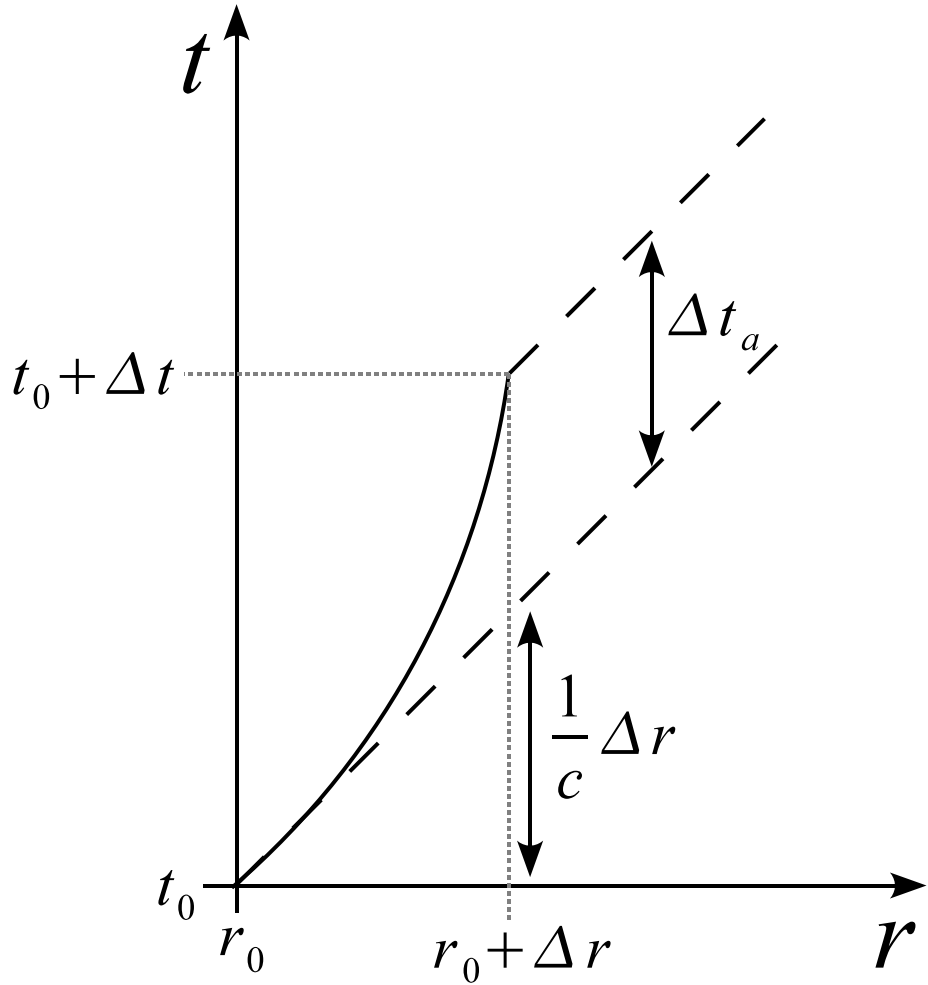} 
\caption{Relation between the arrival time $t_a$ and the laboratory time $t$. Details in Ruffini et al. \cite{lett1,Brasile}.} 
\label{ttasch_new_bn} 
\end{figure} 

The message by Einstein is simply illustrated in Fig. \ref{ttasch_new_bn}. If we consider in an inertial frame a source (solid line) moving with high speed and emitting light signals (dashed lines) along the direction of its motion, a far away observer will measure a delay $\Delta t_a$ between the arrival time of two signals emitted at the origin and after a time interval  $\Delta t$ in the laboratory frame. The real velocity of the source is given by: 
\begin{equation} 
v = \frac{\Delta r}{\Delta t} 
\label{v} 
\end{equation} 
and the apparent velocity is given by: 
\begin{equation} 
v_{app} = \frac{\Delta r}{\Delta t_a}\, , 
\label{vapp} 
\end{equation} 
As pointed out by Einstein the adoption of coordinating light signals simply by their arrival time as in Eq.(\ref{vapp}), without an adequate definition of synchronization, is incorrect and leads to unsurmountable difficulties as well as to apparently ``superluminal'' velocities as soon as motions close to the speed of light are considered.

The use of $\Delta t_a$ as a time coordinate, often tacitly adopted by astronomers, should be done, if at all, with proper care. The relation between $\Delta t_a$ and the correct time parameterization in the laboratory frame has to be taken into account:
\begin{equation} 
\Delta t_a = \Delta t - \frac{\Delta r}{c} = \Delta t - 
\frac{1}{c}\int_{t_\circ}^{t_\circ + \Delta t}{v\left(t'\right) dt'}\, . 
\label{tadef} 
\end{equation} 
In other words, the relation between the arrival time and the laboratory time cannot be done without a knowledge of the speed along the entire world-line of the source. In the case of GRBs, such a worldline starts at the moment of gravitational collapse. It is of course clear that the parameterization in the laboratory frame has to take into account the cosmological redshift $z$ of the source. We then have, at the detector:
\begin{equation}
\Delta t_a^d = \left(1+z\right) \Delta t_a\, .
\label{taddef}
\end{equation}

In the current GRB literature, Eq.(\ref{tadef}) has been systematically neglected by addressing only the afterglow description neglecting the previous history of the source. Often the integral equation has been approximated by a clearly incorrect instantaneous value: 
\begin{equation} 
\Delta t_a \simeq \frac{\Delta t}{2\gamma^2}\, . 
\label{taapp} 
\end{equation}
The attitude has been adopted that it should be possible to consider separately the afterglow part of the GRB phenomenon, without the knowledge of the entire equation of motion of the source. 

This point of view has reached its most extreme expression in the works reviewed by Piran \cite{p99,p00}, where the so-called ``prompt radiation'', lasting on the order of $10^2$ s, is considered as a burst emitted by the prolonged activity of an ``inner engine''. In these models, generally referred to as the ``internal shock model'', the emission of the afterglow is assumed to follow the ``prompt radiation'' phase \cite{rm94,px94,sp97,f99,fcrsyn99}.

As we outline in the following, such an extreme point of view originates from the inability of obtaining the time scale of the ``prompt radiation'' from a burst structure. These authors consequently appeal to the existence of an ``ad hoc'' inner engine in the GRB source to solve this problem.

We show in the following how this difficulty has been overcome in our approach by interpreting the ``prompt radiation'' as an integral part of the afterglow and {\em not} as a burst. This explanation can be reached only through a relativistically correct theoretical description of the entire afterglow (see section \ref{aft}). Within the framework of special relativity we show that it is not possible to describe a GRB phenomenon by disregarding the knowledge of the entire past worldline of the source. We show that at $10^2$ seconds the emission occurs from a region of dimensions of approximately $10^{16}$ cm, well within the region of activity of the afterglow. This point was not appreciated in the current literature due to the neglect of the apparent superluminal effects implied by the use of the ``pathological'' parametrization of the GRB phenomenon by the arrival time of light signals.

An additional difference between our treatment and the ones in the current literature relates to the assumption of the existence of scaling laws in the afterglow phase: the power law dependence of the Lorentz gamma factor on the radial coordinate is usually systematically assumed. From the proper use of the relativistic transformations and by the direct numerical and analytic integration of the special relativistic equations of motion we demonstrate (see section \ref{eqaft}) that no simple power-law relation can be derived for the equations of motion of the system. This situation is not new for workers in relativistic theories: scaling laws exist in the extreme ultrarelativistic regimes and in the Newtonian ones but not in the intermediate fully relativistic regimes (see e.g. Ruffini \cite{r70}).

\subsection{GRBs and general relativity}\label{genrel}

\begin{figure} 
\centering 
\includegraphics[width=\hsize,clip]{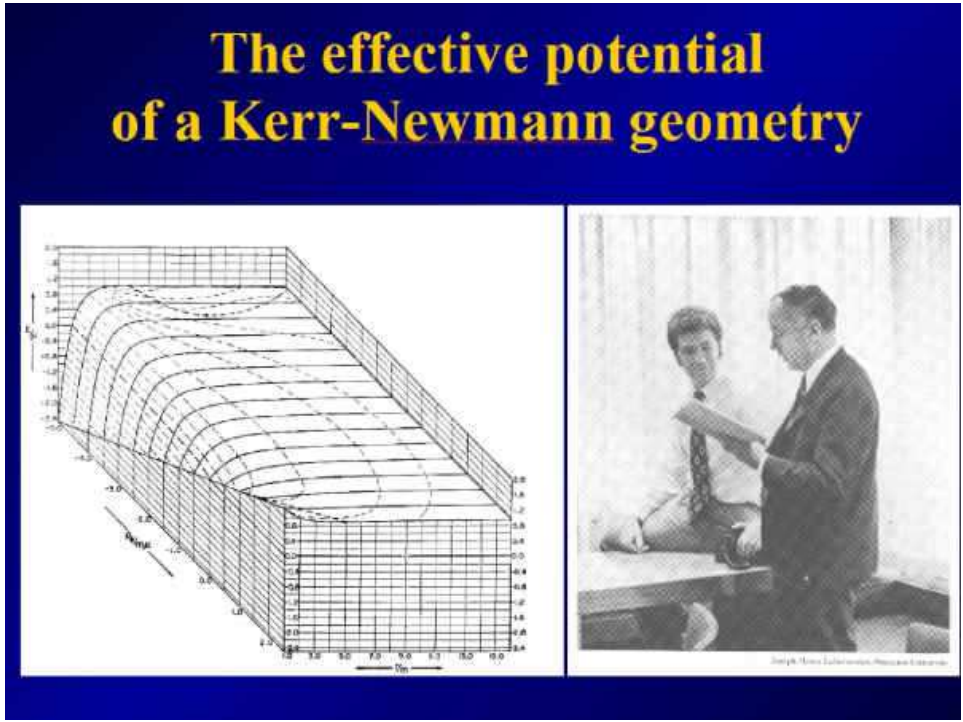} 
\caption{The effective potential corresponding to the circular orbits in the equatorial plane of a black hole is given as a function of the angular momentum of the test particle. This digram was originally derived by Ruffini and Wheeler (right picture). For details see Landau and Lifshitz \cite{ll2} and Rees, Ruffini and Wheeler \cite{rrw}.}
\label{effp} 
\end{figure}

Three of the most important works in the field of general relativity have certainly been the discovery of the Kerr solution \cite{kerr}, its generalization to the charged case (Newman et al. \cite{newman}) and the formulation by Brandon Carter \cite{carter} of the Hamilton-Jacobi equations for a charged test particle in the metric and electromagnetic field of a Kerr-Newman solution (see e.g. Landau and Lifshitz \cite{ll2}). The equations of motion, which are generally second order differential equations, were reduced by Carter to a set of first order differential equations which were then integrated by using an effective potential technique by Ruffini and Wheeler for the Kerr metric (see e.g. Landau and Lifshitz \cite{ll2}) and by Ruffini for the Reissner-Nordstr\"om geometry (Ruffini \cite{r70}, see Fig. \ref{effp}).

All the above mathematical results were essential for understanding the new physics of gravitationally collapsed objects and allowed the publication of a very popular article: ``Introducing the black hole'' (Ruffini and Wheeler \cite{rw71}). In that paper, we advanced the ansatz that the most general black hole is a solution of the Einstein-Maxwell equations, asymptotically flat and with a regular horizon: the Kerr-Newman solution, characterized only by three parameters: the mass $M$, the charge $Q$ and the angular momentum $L$. This ansatz of the ``black hole uniqueness theorem'' still today after thirty years presents challenges to the mathematical aspects of its complete proof (see e.g. Carter \cite{ckf} and Bini et al. \cite{bcjr}). In addition to these mathematical difficulties, in the field of physics this ansatz contains the most profound consequences. The fact that, among all the possible highly nonlinear terms characterizing the gravitationally collapsed objects, only the ones corresponding solely to the Einstein Maxwell equations survive the formation of the horizon has, indeed, extremely profound physical implications. Any departure from such a minimal configuration either collapses on the horizon or is radiated away during the collapse process. This ansatz is crucial in identifying precisely the process of gravitational collapse leading to the formation of the black hole and the emission of GRBs. Indeed, in this specific case, the Born-like nonlinear \cite{b33} term of the Heisenberg-Euler-Schwinger \cite{he35,s51} Lagrangian are radiated away prior to the formation of the horizon of the black hole (see e.g. Ruffini et al. \cite{rvx05}). Only the nonlinearity corresponding solely to the classical Einstein-Maxwell theory is left as the outcome of the gravitational collapse process.

The same effective potential technique (see Landau and Lifshitz \cite{ll2}), which allowed the analysis of circular orbits around the black hole, was crucial in reaching the equally interesting discovery of the reversible and irreversible transformations of black holes by Christodoulou and Ruffini \cite{cr71}, which in turn led to the mass-energy formula of the black hole:
\begin{equation} 
E_{BH}^2 = M^2c^4 = \left(M_{\rm ir}c^2 + \frac{Q^2}{2\rho_+}\right)^2+\frac{L^2c^2}{\rho_+^2}\, ,
\label{em} 
\end{equation} 
with 
\begin{equation} 
\frac{1}{\rho_+^4}\left(\frac{G^2}{c^8}\right)\left(Q^4+4L^2c^2\right)\leq 1\, , 
\label{s1}
\end{equation} 
where 
\begin{equation} 
S=4\pi\rho_+^2=4\pi(r_+^2+\frac{L^2}{c^2M^2})=16\pi\left(\frac{G^2}{c^4}\right) M^2_{\rm ir}\, ,
\label{sa} 
\end{equation} 
is the horizon surface area, $M_{\rm ir}$ is the irreducible mass, $r_{+}$ is the horizon radius and $\rho_+$ is the quasi-spheroidal cylindrical coordinate of the horizon evaluated at the equatorial plane. Extreme black holes satisfy the equality in Eq.(\ref{s1}).

From Eq.(\ref{em}) follows that the total energy of the black hole $E_{BH}$ can be split into three different parts: rest mass, Coulomb energy and rotational energy. In principle both Coulomb energy and rotational energy can be extracted from the black hole (Christodoulou and Ruffini \cite{cr71}). The maximum extractable rotational energy is 29\% and the maximum extractable Coulomb energy is 50\% of the total energy, as clearly follows from the upper limit for the existence of a black hole, given by Eq.(\ref{s1}). We refer in the following to both these extractable energies as the blackholic energy.

The existence of the black hole and the basic correctness of the circular orbits has been proven by the observations of Cygnus-X1 (see e.g. Giacconi and Ruffini \cite{gr78}). However, in binary X-ray sources, the black hole uniquely acts passively by generating the deep potential well in which the accretion process occurs. It has become tantalizing to look for astrophysical objects in order to verify the other fundamental prediction of general relativity that the blackholic energy is the largest energy extractable from any physical object.

As we shall see in the next section, the feasibility of the extraction of the blackholic energy has been made possible by the quantum processes of creating, out of classical fields, a plasma of electron-positron pairs in the field of black holes. The manifestation of such process of energy extraction from the black hole is astrophysically manifested by the occurrence of GRBs.

\subsection{GRBs and quantum electro-dynamics} 

That a static electromagnetic field stronger than a critical value: 
\begin{equation} 
E_c = \frac{m_e^2c^3}{\hbar e} 
\label{ec} 
\end{equation} 
can polarize the vacuum and create electron-positron pairs was clearly evidenced by Heisenberg and Euler \cite{he35}. The major effort in verifying the correctness of this theoretical prediction has been directed in the analysis of heavy ion collisions (see Ruffini et al. \cite{rvx05} and references therein). From an order-of-magnitude estimate, it appears that around a nucleus with a charge: 
\begin{equation} 
Z_c \simeq \frac{\hbar c}{e^2} \simeq 137 
\label{zc} 
\end{equation} 
the electric field can be stronger than the electric field polarizing the vacuum. A more accurate detailed analysis taking into account the bound states levels around a nucleus brings to a value of
\begin{equation} 
Z_c \simeq 173
\label{zc2} 
\end{equation}  
for the nuclear charge leading to the existence of a critical field. From the Heisenberg uncertainty principle it follows that, in order to create a pair, the existence of the critical field should last a time
\begin{equation} 
\Delta t \sim \frac{\hbar}{m_e c^2} \simeq 10^{-18}\, \mathrm{s}\, ,
\label{dt} 
\end{equation} 
which is much longer then the typical confinement time in heavy ion collisions which is 
\begin{equation} 
\Delta t \sim \frac{\hbar}{m_p c^2} \simeq 10^{-21}\, \mathrm{s}\, .
\label{dt2} 
\end{equation} 
This is certainly a reason why no evidence for pair creation in heavy ion collisions has been obtained although remarkable effort has been spent in various accelerators worldwide. Similar experiments involving laser beams encounter analogous difficulties (see e.g. Ruffini et al. \cite{rvx05} and references therein).

The alternative idea was advanced in 1975 \cite{dr75} that the critical field condition given in Eq.(\ref{ec}) could be reached easily, and for a time much larger than the one given by Eq.(\ref{dt}), in the field of a Kerr-Newman black hole in a range of masses $3.2M_\odot \le M_{BH} \le 7.2\times 10^6M_\odot$. In that paper we have generalized to the curved Kerr-Newman geometry the fundamental theoretical framework developed in Minkowski space by Heisenberg-Euler \cite{he35} and Schwinger \cite{s51}. This result was made possible by the work on the structure of the Kerr-Newman spacetime previously done by Carter \cite{carter} and by the remarkable mathematical craftsmanship of Thibault Damour then working with me as a post-doc in Princeton.

The maximum energy extractable in such a process of creating a vast amount of electron-positron pairs around a black hole is given by:
\begin{equation} 
E_{max} = 1.8\times 10^{54} \left(M_{BH}/M_\odot\right)\, \mathrm{erg}\, \mathrm{.} 
\label{emax} 
\end{equation} 
We concluded in that paper that such a process ``naturally leads to a most simple model for the explanation of the recently discovered $\gamma$-rays bursts''.

At that time, GRBs had not yet been optically identified and nothing was known about their distance and consequently about their energetics. Literally thousands of theories existed in order to explain them and it was impossible to establish a rational dialogue with such an enormous number of alternative theories. We did not pursue further our model until the results of the BeppoSAX mission, which clearly pointed to the cosmological origin of GRBs, implying for the typical magnitude of their energy precisely the one predicted by our model.

It is interesting that the idea of using an electron-positron plasma as a basis of a GRB model was independently introduced years later in a set of papers by Cavallo and Rees \cite{cr78}, Cavallo and Horstman \cite{ch81} and Horstman and Cavallo \cite{hc83}. These authors did not address the issue of the physical origin of their energy source. They reach their conclusions considering the pair creation and annihilation process occurring in the confinement of a large amount of energy in a region of dimension $\sim 10$ km typical of a neutron star. No relation to the physics of black holes nor to the energy extraction process from a black hole was envisaged in their interesting considerations, mainly directed to the study of the opacity and the consequent dynamics of such an electron-positron plasma.

\begin{figure} 
\centering 
\includegraphics[width=\hsize,clip]{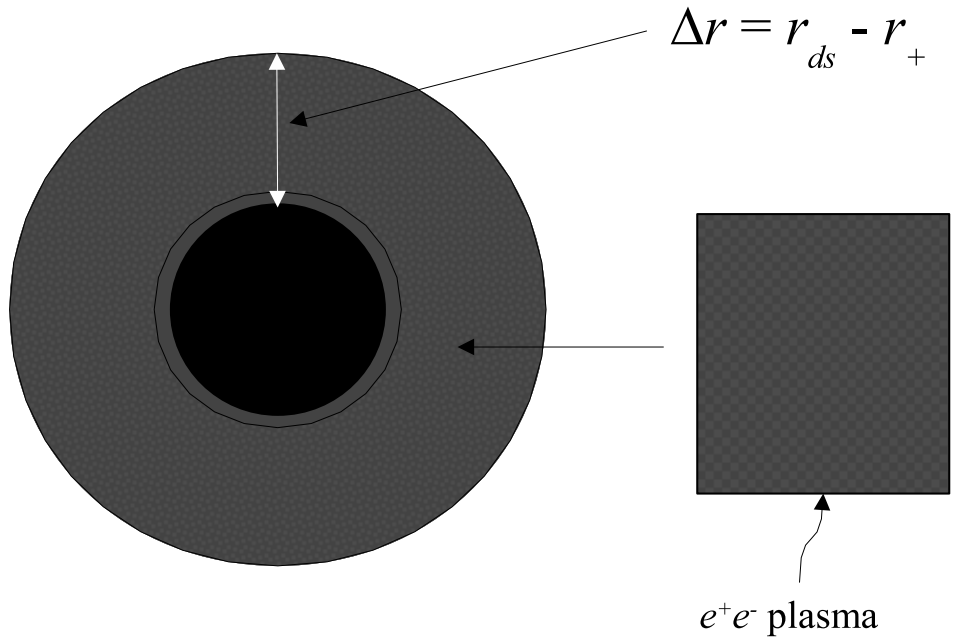} 
\caption{The dyadosphere is comprised between the horizon radius and the radius of the dyadosphere. All this region is filled with electron-positron pairs and photons in thermal equilibrium. Details in Ruffini \cite{rukyoto}, Preparata et al. \cite{prx98}, Ruffini et al. \cite{rvx03a}.} 
\label{dya} 
\end{figure} 

After the discovery of the afterglows and the optical identification of GRBs at cosmological distances, implying exactly the energetics predicted in Eq.(\ref{emax}), we returned to the analysis of the vacuum polarization process around a black hole and precisely identified the region around the black hole in which the vacuum polarization process and the consequent creation of electron-positron pairs occur. We defined this region, using the Greek name dyad for pairs ($\delta\upsilon\alpha\varsigma$, $\delta\upsilon\alpha\delta o \varsigma$), to be the ``dyadosphere'' of the black hole, bounded  by the black hole horizon and the dyadosphere radius $r_{ds}$ given by (see Ruffini \cite{rukyoto}, Preparata et al. \cite{prx98} and Fig.\ref{dya}):
\begin{equation} 
r_{ds}=\left(\frac{\hbar}{mc}\right)^\frac{1}{2}\left(\frac{GM}{ 
c^2}\right)^\frac{1}{2} \left(\frac{m_{\rm p}}{m}\right)^\frac{1}{2}\left(\frac{e}{q_{\rm p}}\right)^\frac{1}{2}\left(\frac{Q}{\sqrt{G}M}\right)^\frac{1}{2}=1.12\cdot 10^8\sqrt{\mu\xi} \, {\rm cm}, 
\label{rc} 
\end{equation} 
where we have introduced the dimensionless mass and charge parameters $\mu={M_{BH}/M_{\odot}}$, $\xi={Q/(M_{BH}\sqrt{G})}\le 1$. 

The analysis of the dyadosphere was developed, at that time, around an already formed black hole. In recent months we have been developing the dynamical formation of the black hole and correspondingly of the dyadosphere during the process of gravitational collapse, reaching some specific signatures which may be detectable in the structure of the short and long GRBs (Cherubini et al. \cite{crv02}, Ruffini and Vitagliano \cite{rv02a,rv02b}, Ruffini et al. \cite{rvx03a,rvx03b,rfvx05}).

\section{The dynamical phases following the dyadosphere formation}

Many details of this topic have been presented in great details in Ruffini et al. \cite{Brasile}.

After the vacuum polarization process around a black hole, one of the topics of the greatest scientific interest is the analysis of the dynamics of the electron-positron plasma formed in the dyadosphere. This issue was addressed by us in a very effective collaboration with Jim Wilson at Livermore. The numerical simulations of this problem were developed at Livermore, while the semi-analytic approach was developed in Rome (Ruffini et al. \cite{rswx99}).

The corresponding treatment in the framework of the Cavallo et al. analysis was performed by Piran et al. \cite{psn93} also using a numerical approach, by Bisnovaty-Kogan and Murzina \cite{bm95} using an analytic approach and by M\'esz\'aros, Laguna and Rees \cite{mlr93} using a numerical and semi-analytic approach.

Although some analogies exists between these treatments, they are significantly different in the theoretical details and in the final results. Since the final result of the GRB model is extremely sensitive to any departure from the correct treatment, it is indeed very important to detect at every step the appearance of possible fatal errors.

A conclusion common to all these treatments is that the electron-positron plasma is initially optically thick and expands till transparency reaching very high values of the Lorentz gamma 
factor. A second point, which is common, is the discovery of a new clear feature: the plasma shell expands but the Lorentz contraction is such that its width in the laboratory frame appears to be constant.

\begin{figure}
\centering
\includegraphics[width=\hsize,clip]{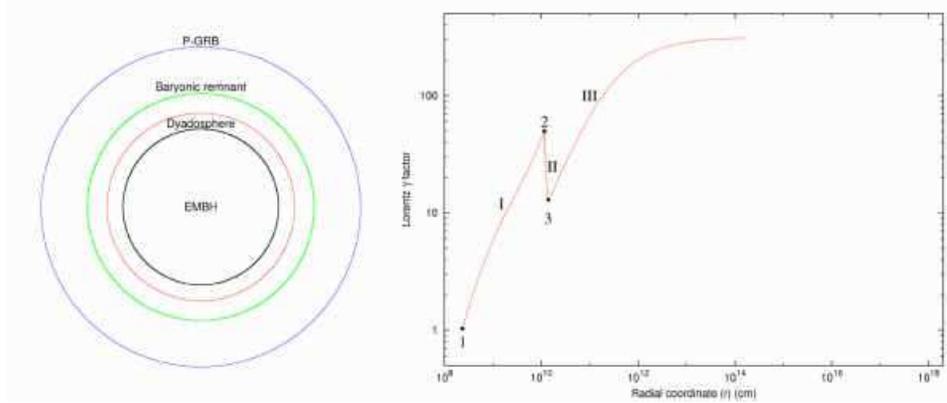}
\caption{The optically thick phase of our model are qualitatively represented in this diagram. There are clearly recognizable 1) the PEM pulse phase, 2) the impact on the baryonic remnant, 3) the PEMB pulse phase and the final approach to transparency with the emission of the P-GRB (see Fig. \ref{cip3}). Details in Ruffini et al. \cite{Brasile}.}
\label{cip2}
\end{figure}

There is however a major difference between our approach and the ones of Piran, M\'esz\'aros and Rees, in that the dyadosphere is assumed by us to be filled uniquely with an electron-positron plasma. Such a plasma expands in substantial agreement with the results presented in the work of Bisnovati-Kogan and Murzina \cite{bm95}. In our model the pulse of electron-positron pairs and photons (PEM Pulse, see Ruffini et al. \cite{rswx99}) evolves and at a radius on the order of $10^{10}$ cm it encounters the remnant of the star progenitor of the newly formed black hole. The PEM pulse is then loaded with baryons. A new pulse is  formed of electron-positron-photons and baryons (PEMB Pulse, see Ruffini et al. \cite{rswx00}) which expands all the way until transparency is reached. At transparency the emitted photons give origin to what we define as the Proper-GRB (see Ruffini et al. \cite{lett2} and Fig. \ref{cip2}).

In our approach, the baryon loading is measured by a dimensionless quantity
\begin{equation}
B = \frac{M_B c^2}{E_{dya}}\, ,
\label{Bdef}
\end{equation}
which gives direct information about the mass $M_B$ of the remnant. The corresponding treatment done by Piran and collaborators (Shemi \& Piran \cite{sp90}, Piran et al. \cite{psn93}) and by M\'esz\'aros, Laguna and Rees \cite{mlr93} differs in one important respect: the baryonic loading is assumed to occur since the beginning of the electron-positron pair formation and no relation to the mass of the remnant of the collapsed progenitor star is attributed to it.

A marked difference also exists between our description of the rate equation for the electron-positron pairs and the ones by those authors. While our results are comparable with the ones obtained by Piran under the same initial conditions, the set of approximations adopted by M\'esz\'aros, Laguna and Rees \cite{mlr93} appears to be too radical and leads to very different results violating energy and momentum conservation (see Bianco et al. \cite{bfrvx}).

From our analysis (Ruffini et al. \cite{rswx00}) it also becomes clear that such expanding dynamical evolution can only occur for values of $B < 10^{-2}$. This prediction, as we will show shortly in the three GRB sources considered here, is very satisfactorily confirmed by observations.

From the value of the $B$ parameter, related to the mass of the remnant, it therefore follows that the collapse to a black hole leading to a GRB is drastically different from the collapse to a neutron star. While in the case of a neutron star collapse a very large amount of matter is expelled, in many instances well above the mass of the neutron star itself, in the case of black holes leading to a GRB only a very small fraction of the initial mass ($\sim 10^{-2}$ or less) is expelled. The collapse to a black hole giving rise to a GRB appears to be much smoother than any collapse process considered until today: almost 99.9\% of the star has to be collapsing simultaneously!

\begin{figure} 
\centering 
\includegraphics[width=\hsize,clip]{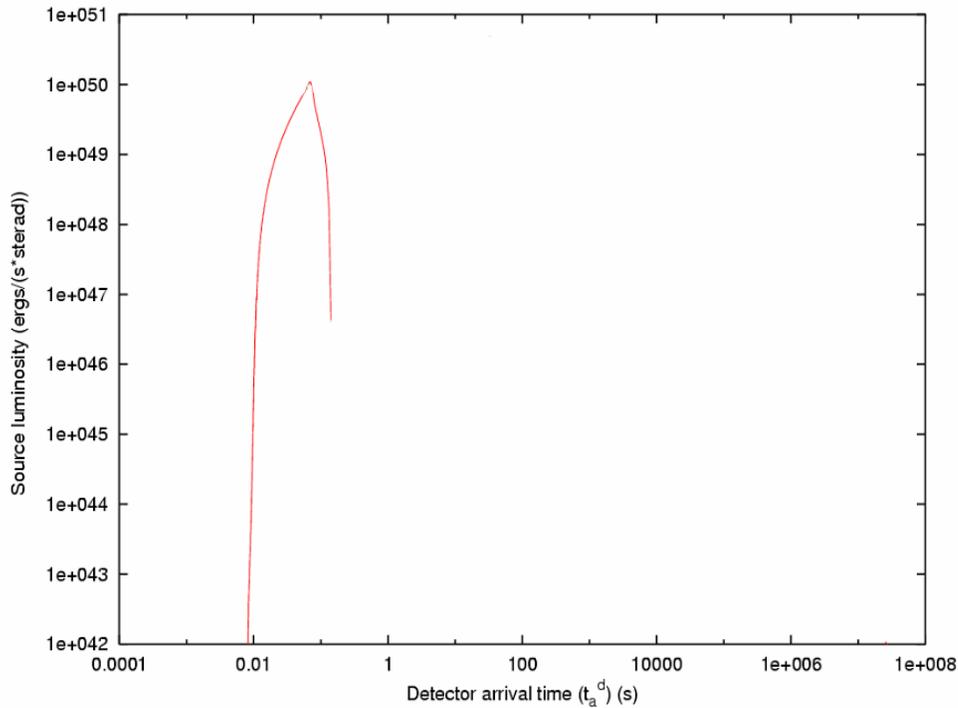} 
\caption{The P-GRB emitted at the transparency point at a time of arrival $t_a^d$ which has been computed following the prescriptions of Eq.(\ref{tadef}). Details in Ruffini et al. \cite{lett2,Brasile}.} 
\label{cip3} 
\end{figure}

We summarize in Figs. \ref{cip2}--\ref{cip3} the optically thick phase of GRBs in our model: we start from a given dyadosphere of energy $E_{dya}$; the pair-electromagnetic pulse (PEM pulse) self-accelerates outward typically reaching Lorentz gamma factors $\gamma \sim 200$ at $r \sim 10^{10}$ cm; at this point the collision of the PEM pulse with the remnant of the progenitor star occurs with an abrupt decrease in the value of the Lorentz gamma factor; a new pair-electromagnetic-baryon pulse (PEMB pulse) is formed which self-accelerates outward until the system becomes transparent.

The photon emission at this transparency point is the Proper-GRB (P-GRB). An accelerated beam of baryons with an initial Lorentz gamma factor $\gamma_\circ$ starts to interact with the interstellar medium at typical distances from the black hole of $r_\circ \sim 10^{14}$ cm and at a photon arrival time at the detector on the Earth surface of $t_a^d \sim 0.1$ s. These values determine the initial conditions of the afterglow.

\section{The description of the afterglow}\label{aft}

\begin{figure} 
\centering 
\includegraphics[width=\hsize,clip]{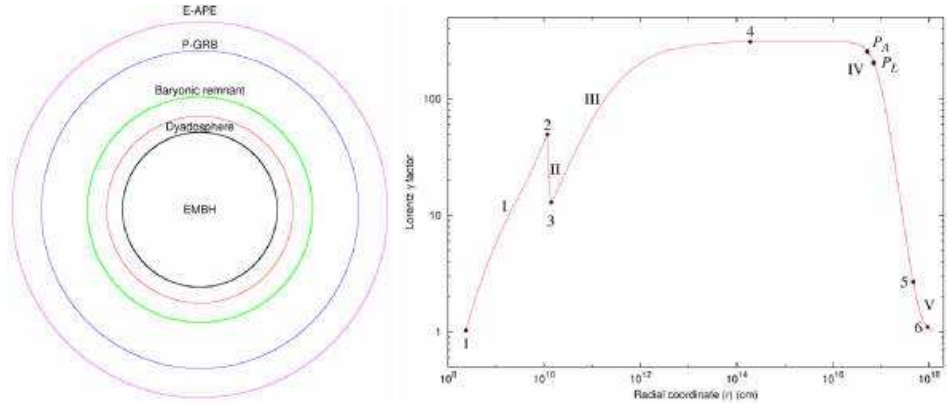} 
\caption{The GRB afterglow phase is here represented together with the optically thick phase (see Fig. \ref{cip2}). The value of the Lorentz gamma factor is here given from the transparency point all the way to the ultrarelativisitc, relativistic and non relativistic regimes. Details in Ruffini et al. \cite{Brasile}.}
\label{cip_tot} 
\end{figure}

After reaching transparency and the emission of the P-GRB, the accelerated baryonic matter (the ABM pulse) interacts with the interstellar medium (ISM) and gives rise to the afterglow (see Fig. \ref{cip_tot}). Also in the descriptions of this last phase many differences exist between our treatment and the other ones in the current literature. 

\subsection{The initial value problem}

The initial conditions of the afterglow era are determined at the end of the optically thick era when the P-GRB is emitted. As recalled in the last section, the transparency condition is determined by a time of arrival $t_a^d$, a value of the gamma Lorentz factor $\gamma_\circ$, a value of the radial coordinate $r_\circ$, an amount of baryonic matter $M_B$ which are only functions of the two parameters $E_{dya}$ and $B$ (see Eq.(\ref{Bdef})). It is appropriate here to emphasize again that, in order to have the expansion leading to an observed GRB, one must have $B < 10^{-2}$.

This connection to the optically thick era is missing in the current approach in the literature which attributes the origin of the ``prompt radiation'' to an unspecified inner engine activity (see Piran \cite{p99} and references therein). The initial conditions at the beginning of the afterglow era are obtained by a best fit of the later parts of the afterglow. This approach is quite unsatisfactory since, as we will explicitly show, the theoretical treatments currently adopted in the description of the afterglow are not correct. The fit using an incorrect theoretical treatment leads necessarily to the wrong conclusions as well as, in turn, to the determination of incorrect initial conditions.

\subsection{The equations of the afterglow dynamics}\label{eqaft}

Let us first summarize the commonalities between our approach and the ones in the current literature. In both cases (see Piran \cite{p99}, Chiang \& Dermer \cite{cd99} and Ruffini et al. \cite{Brasile}) a thin shell approximation is used to describe the collision between the ABM pulse and the ISM: 
\begin{subequations}\label{Taub_Eq} 
\begin{eqnarray} 
dE_{\mathrm{int}} &=& \left(\gamma - 1\right) dM_{\mathrm{ism}} c^2 
\label{Eint}\, ,\\ 
d\gamma &=& - \textstyle\frac{{\gamma}^2 - 1}{M} dM_{\mathrm{ism}}\, , 
\label{gammadecel}\\ 
dM &=& 
\textstyle\frac{1-\varepsilon}{c^2}dE_{\mathrm{int}}+dM_\mathrm{ism}\, 
,\label{dm}\\ 
dM_\mathrm{ism} &=& 4\pi m_p n_\mathrm{ism} r^2 dr \, , \label{dmism} 
\end{eqnarray} 
\end{subequations} 
where $E_{\mathrm{int}}$, $\gamma$ and $M$ are respectively the internal energy, the Lorentz factor and the mass-energy of the expanding pulse, $n_\mathrm{ism}$ is the ISM number density which is assumed to be constant, $m_p$ is the proton mass, $\varepsilon$ is the emitted fraction of the energy developed in the collision with the ISM and $M_\mathrm{ism}$ is the amount of ISM mass swept up within the radius $r$: $M_\mathrm{ism}=(4/3)\pi(r^3-{r_\circ}^3)m_pn_\mathrm{ism}$, where $r_\circ$ is the starting radius of the shock front. In general, an additional condition is needed in order to determine $\varepsilon$ as a function of the radial coordinate. In the following, $\varepsilon$ is 
assumed to be constant and such an approximation appears to be correct in the GRB context.

In both our work and in the current literature (see Piran \cite{p99}, Chiang \& Dermer \cite{cd99} and Ruffini et al. \cite{Brasile}) a first integral of these equations has been found, leading to expressions for the Lorentz gamma factor as a function of the radial coordinate. In the ``fully adiabatic condition'' (i.e. $\varepsilon = 0$) we have:
\begin{equation} 
\gamma^2=\frac{\gamma_\circ^2+2\gamma_\circ\left(M_\mathrm{ism}/M_B\right) 
+\left(M_\mathrm{ism}/M_B\right)^2}{1+2\gamma_\circ\left(M_\mathrm{ism}/M_B\right)+\left(M_\mathrm{ism}/M_B\right)^2}\, , 
\label{gamma_ad} 
\end{equation} 
while in the ``fully radiative condition'' (i.e. $\varepsilon = 1$) we have: 
\begin{equation} 
\gamma=\frac{1+\left(M_\mathrm{ism}/M_B\right)\left(1+\gamma_\circ^{-1}\right)\left[1+\left(1/2\right)\left(M_\mathrm{ism}/M_B\right)\right]}{\gamma_\circ^{-1}+\left(M_\mathrm{ism}/M_B\right)\left(1+\gamma_\circ^{-1}
\right)\left[1+\left(1/2\right)\left(M_\mathrm{ism}/M_B\right)\right]}\, , 
\label{gamma_rad} 
\end{equation} 
where $\gamma_\circ$ and $M_B$ are respectively the values of the Lorentz gamma factor and of the mass of the accelerated baryons at the beginning of the afterglow phase and $r_\circ$ is the value of the radius $r$ at the beginning of the afterglow phase. 

A major difference between our treatment and the other ones in the current literature is that we have integrated the above equations analytically, obtaining the explicit analytic form of the equations of motion for the expanding shell in the afterglow for a constant ISM density. For the fully radiative case we have explicitly integrated the differential equation for $r\left(t\right)$ in Eq.\eqref{gamma_rad}, recalling that $\gamma^{-2}=1-\left[dr/\left(cdt\right)\right]^2$, where $t$ is the time in the laboratory reference frame. We have then obtained a new explicit analytic solution of the equations of motion for the relativistic shell in the entire range from the ultra-relativistic to the non-relativistic regimes:
\begin{equation} 
\begin{split} 
& t = \tfrac{M_B  - m_i^\circ}{2c\sqrt C }\left( {r - r_\circ } \right) 
+ \tfrac{{r_\circ \sqrt C }}{{12cm_i^\circ A^2 }} \ln \left\{ 
{\tfrac{{\left[ {A + \left(r/r_\circ\right)} \right]^3 \left(A^3  + 
1\right)}}{{\left[A^3  + \left( r/r_\circ \right)^3\right] \left( {A + 1} 
\right)^3}}} \right\} - \tfrac{m_i^\circ r_\circ }{8c\sqrt C }\\ 
& + t_\circ + \tfrac{m_i^\circ r_\circ }{8c\sqrt C } \left( 
{\tfrac{r}{{r_\circ }}} \right)^4 + \tfrac{{r_\circ \sqrt{3C}}}{{6 c 
m_i^\circ A^2 }} \left[\arctan \tfrac{{2\left(r/r_\circ\right) - 
A}}{{A\sqrt 3 }} - \arctan \tfrac{{2 - A}}{{A\sqrt 3 }}\right] 
\end{split} 
\label{analsol} 
\end{equation} 
where $A=\sqrt[3]{\left(M_B-m_i^\circ\right)/m_i^\circ}$, $C={M_B}^2(\gamma_\circ-1)/(\gamma_\circ +1)$ and $m_i^\circ = \left(4/3\right)\pi m_p n_{\mathrm{ism}} r_\circ^3$.

Correspondingly, in the adiabatic case we have: 
\begin{equation} 
t = \left(\gamma_\circ-\tfrac{m_i^\circ}{M_B}\right)\tfrac{r-r_\circ}{c\sqrt{\gamma_\circ^2-1}} 
+ \tfrac{m_i^\circ}{4M_Br_\circ^3}\tfrac{r^4-r_\circ^4}{c\sqrt{\gamma_\circ^2-1}} 
+ t_\circ\, . 
\label{analsol_ad} 
\end{equation} 

In the current literature, following Blandford and McKee \cite{bm76}, a so-called ``ultrarelativistic'' approximation $\gamma_\circ \gg \gamma \gg 1$ has been widely adopted by many authors to solve Eqs.\eqref{Taub_Eq} (see e.g. Sari \cite{s97,s98}, Waxman \cite{w97}, Rees \& M\'esz\'aros \cite{rm98}, Granot et al. \cite{gps99}, Panaitescu \& M\'esz\'aros \cite{pm98c}, Piran \cite{p99}, Gruzinov \& Waxman \cite{gw99}, van Paradijs et al. \cite{vpkw00}, M\'esz\'aros \cite{m02} and references therein). This leads to simple constant-index power-law relations: 
\begin{subequations}\label{gr0-rct} 
\begin{equation} 
\gamma\propto r^{-a}\, , 
\label{gr0} 
\end{equation} 
with $a=3$ in the fully radiative case and $a=3/2$ in the fully adiabatic case. This simple relation is in stark contrast to the complexity of Eq.\eqref{gamma_rad} and Eq.\eqref{gamma_ad} respectively. In the same spirit, instead of Eq.\eqref{analsol} and Eq.\eqref{analsol_ad}, some authors have assumed the following much simpler approximation for the relation between the time and the radial coordinate of the expanding shell, both in the fully radiative and in the fully adiabatic cases:
\begin{equation} 
ct=r\, , 
\label{rct} 
\end{equation} 
while others, like e.g. Panaitescu \& M\'esz\'aros \cite{pm98c}, have integrated the approximate Eq.\eqref{gr0}, obtaining: 
\begin{equation} 
ct=r\left[1+\left(4a+2\right)^{-1}\gamma^{-2}\left(r\right)\right]\, . 
\label{t_app_pm98c} 
\end{equation} 
\end{subequations} 
Again, it is appropriate here to emphasize the stark contrast between Eqs.\eqref{rct},\eqref{t_app_pm98c} and the exact analytic solutions of Eqs.\eqref{Taub_Eq}, expressed in Eqs.\eqref{analsol},\eqref{analsol_ad}.

\subsection{The equitemporal surfaces (EQTSs)}\label{eqts}

As pointed out long ago by Couderc \cite{c39}, in all relativistic expansion the crucial geometrical quantities with respect to a physical observer are the ``equitemporal surfaces'' (EQTSs), namely the locus of source points of the signals arriving at the observer at the same time.

\begin{figure}
\centering
\includegraphics[width=\hsize,clip]{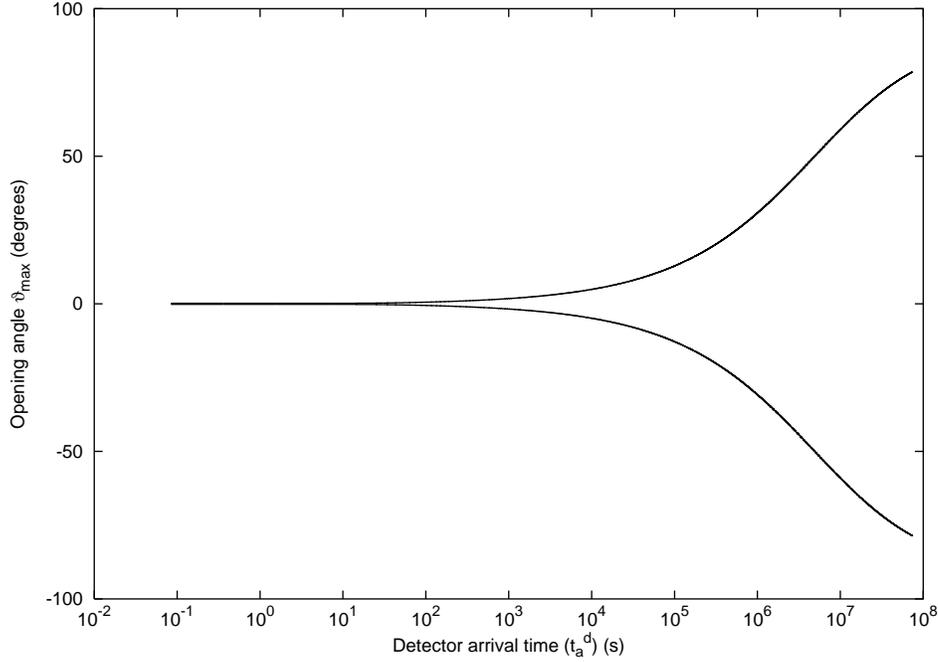}
\caption{Not all values of $\vartheta$ are allowed. Only photons emitted at an angle such that $\cos\vartheta \ge \left(v/c\right)$ can be viewed by the observer. Thus the maximum allowed $\vartheta$ value $\vartheta_{max}$ corresponds to $\cos\vartheta_{max} = (v/c)$. In this figure we show $\vartheta_{max}$ (i.e. the angular amplitude of the visible area of the ABM pulse) in degrees as a function of the arrival time at the detector for the photons emitted along the line of sight (see text). In the earliest GRB phases $v\sim c$ and so $\vartheta_{max}\sim 0$. On the other hand, in the latest phases of the afterglow the ABM pulse velocity decreases and $\vartheta_{max}$ tends to the maximum possible value, i.e. $90^\circ$. Details in Ruffini et al. \cite{rbcfx02_letter,Brasile}}
\label{openang}
\end{figure}

\begin{figure}
\centering
\includegraphics[width=\hsize,clip]{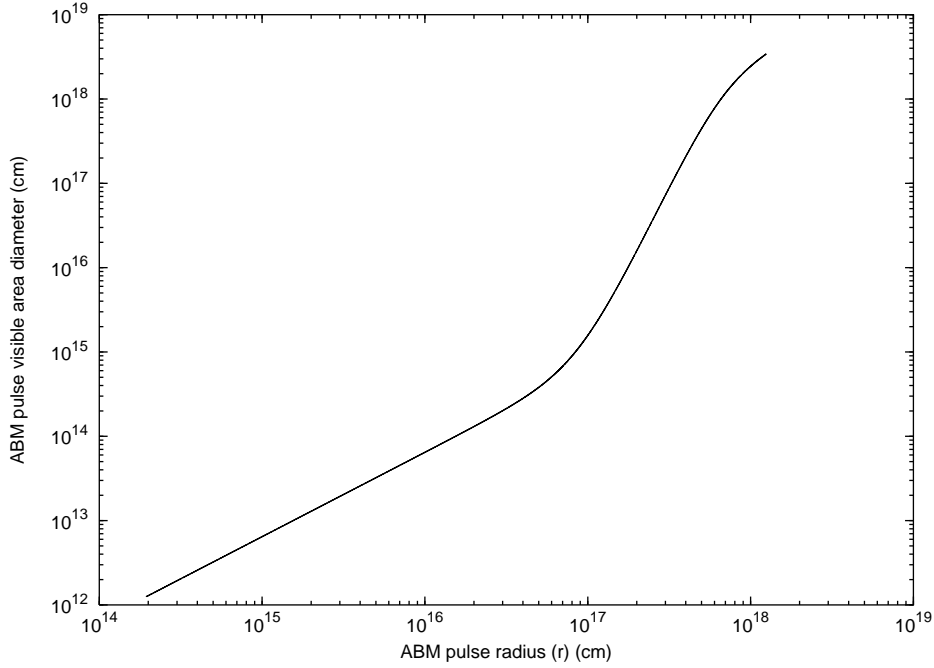}
\caption{The diameter of the visible area is represented as a function of the ABM pulse radius. In the earliest expansion phases ($\gamma\sim 310$) $\vartheta_{max}$ is very small (see left pane and Fig.~\ref{opening}), so the visible area is just a small fraction of the total ABM pulse surface. On the other hand, in the final expansion phases $\vartheta_{max} \to 90^\circ$ and almost all the ABM pulse surface becomes visible. Details in Ruffini et al. \cite{rbcfx02_letter,Brasile}}
\label{opensrad}
\end{figure}

\begin{figure}
\centering
\includegraphics[width=8.0cm,clip]{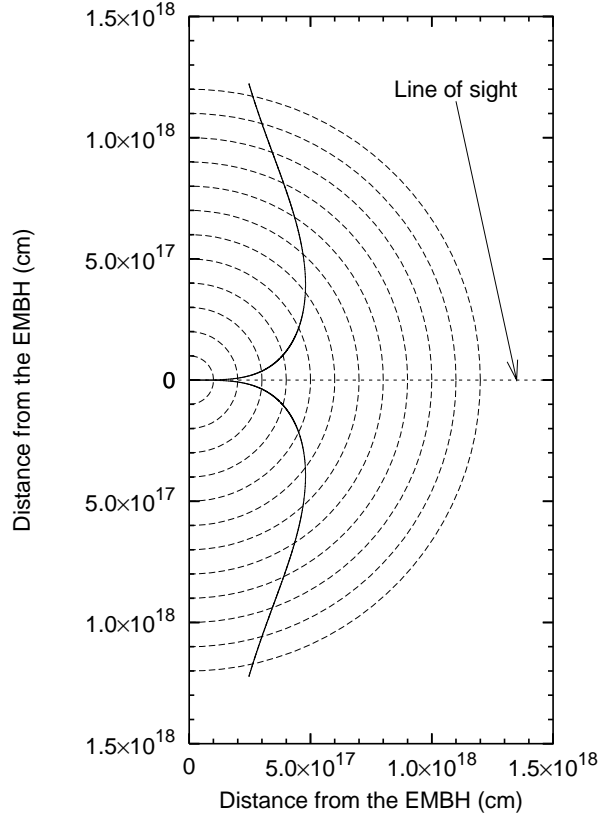}
\caption{This figure shows the temporal evolution of the visible area of the ABM pulse. The dashed half-circles are the expanding ABM pulse at radii corresponding to different laboratory times. The black curve marks the boundary of the visible region. The black hole is located at position (0,0) in this plot. Again, in the earliest GRB phases the visible region is squeezed along the line of sight, while in the final part of the afterglow phase almost all the emitted photons reach the observer. This time evolution of the visible area is crucial to the explanation of the GRB temporal structure. Details in Ruffini et al. \cite{rbcfx02_letter,Brasile}}
\label{opening}
\end{figure}

For a relativistically expanding spherically symmetric source the EQTSs are surfaces of revolution about the line of sight. The general expression for their profile, in the form $\vartheta = \vartheta(r)$, corresponding to an arrival time $t_a$ of the photons at the detector, can be obtained from (see e.g. Ruffini et al. \cite{Brasile}, Bianco and Ruffini \cite{EQTS_ApJL,EQTS_ApJL2} and Figs. \ref{openang}--\ref{opening}):
\begin{equation} 
ct_a = ct\left(r\right) - r\cos \vartheta  + r^\star\, , 
\label{ta_g} 
\end{equation} 
where $r^\star$ is the initial size of the expanding source, $\vartheta$ is the angle between the radial expansion velocity of a point on its surface and the line of sight, and $t = t(r)$ is its equation of motion, expressed in the laboratory frame, obtained by the integration of Eqs.(\ref{Taub_Eq}). From the definition of the Lorentz gamma factor $\gamma^{-2}=1-(dr/cdt)^2$, we have in fact:
\begin{equation} 
ct\left(r\right)=\int_0^r\left[1-\gamma^{-2}\left(r'\right)\right]^{-1/2}dr'\, , 
\label{tdir} 
\end{equation} 
where $\gamma(r)$ comes from the integration of Eqs.(\ref{Taub_Eq}). 

\begin{figure}
\centering
\includegraphics[width=\hsize,clip]{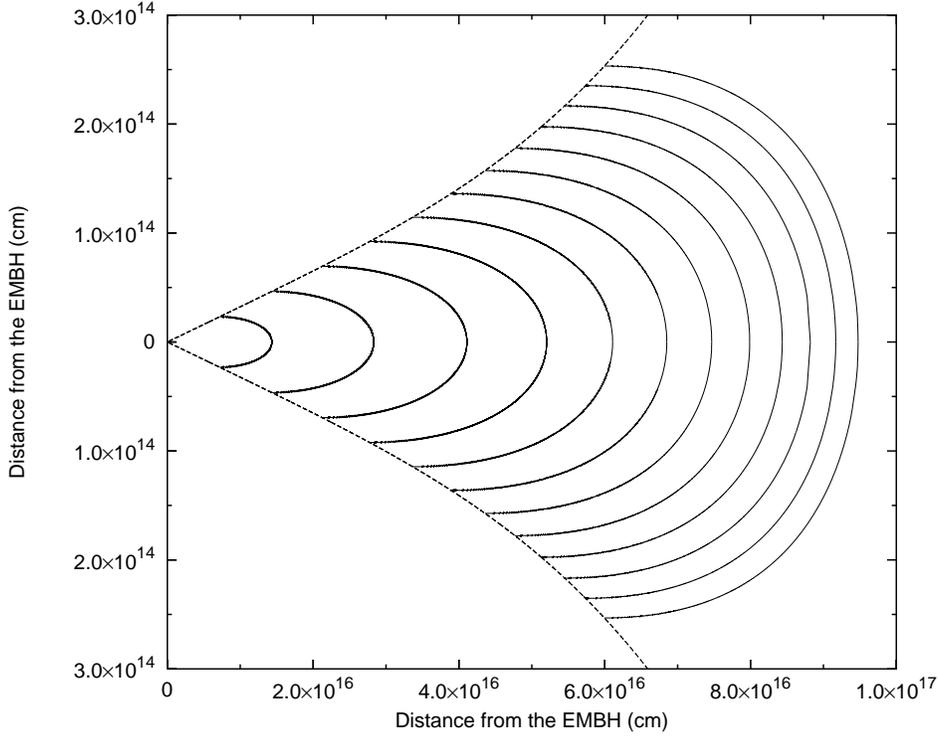}
\caption{Due to the extremely high and extremely varying Lorentz gamma factor, photons reaching the detector on the Earth at the same arrival time are actually emitted at very different times and positions. We represent here the surfaces of photon emission corresponding to selected values of the photon arrival time at the detector: the {\em equitemporal surfaces} (EQTS). Such surfaces differ from the ellipsoids described by Rees in the context of the expanding radio sources with typical Lorentz factor $\gamma\sim 4$ and constant. In fact, in GRB~991216 the Lorentz gamma factor ranges from $310$ to $1$. The EQTSs represented here (solid lines) correspond respectively to values of the arrival time ranging from $5\, s$ (the smallest surface on the left of the plot) to $60\, s$ (the largest one on the right). Each surface differs from the previous one by $5\, s$. To each EQTS contributes emission processes occurring at different values of the Lorentz gamma factor. The dashed lines are the boundaries of the  visible area of the ABM pulse and the black hole is located at position $(0,0)$ in this plot. Note the different scales on the two axes, indicating the very high EQTS ``effective eccentricity''. The time interval from $5\, {\rm s}$ to $60\, {\rm s}$ has been chosen to encompass the E-APE emission, ranging from $\gamma=308.8$ to $\gamma=56.84$. Details in Ruffini et al. \cite{rbcfx02_letter,Brasile}}
\label{ETSNCF}
\end{figure}

\begin{figure} 
\includegraphics[width=\hsize,clip]{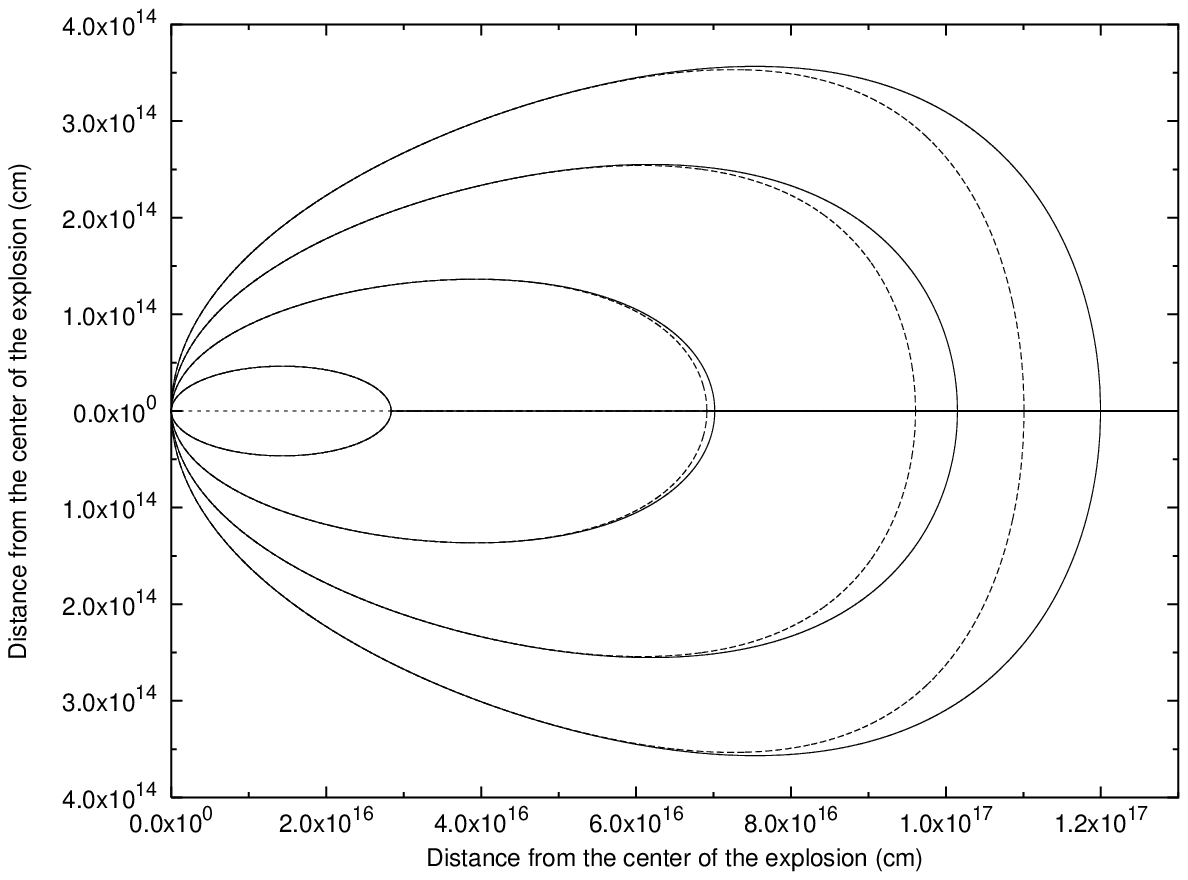} 
\includegraphics[width=\hsize,clip]{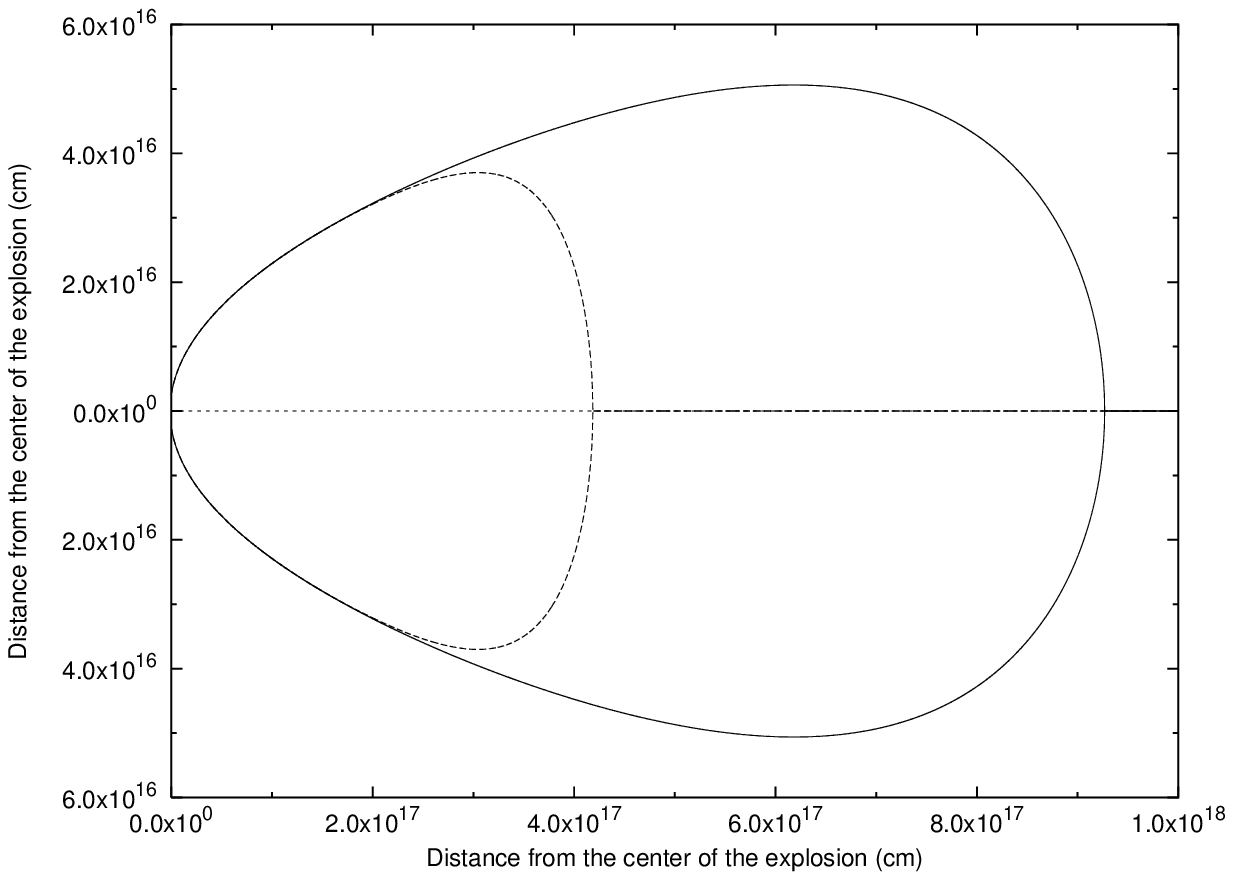} 
\caption{Comparison between EQTSs in the adiabatic regime (solid lines) and in the fully radiative regime (dashed lines). The upper plot shows the EQTSs for $t_a=5$ s, $t_a=15$ s, $t_a=30$ s and $t_a=45$ s, respectively from the inner to the outer one. The lower plot shows the EQTS at an arrival time of 2 days. Details in Bianco and Ruffini \cite{EQTS_ApJL2}.}
\label{eqts_comp} 
\end{figure}

We have obtained the expressions in the adiabatic case and in the fully radiative cases respectively (see Bianco and Ruffini \cite{EQTS_ApJL2}):
\begin{equation} 
\begin{split} 
\cos\vartheta & = 
\frac{m_i^\circ}{4M_B\sqrt{\gamma_\circ^2-1}}\left[\left(\frac{r}{r_\circ}\right)^3 
  - \frac{r_\circ}{r}\right] + \frac{ct_\circ}{r} \\[6pt] & - 
\frac{ct_a}{r} + \frac{r^\star}{r} - 
\frac{\gamma_\circ-\left(m_i^\circ/M_B\right)}{\sqrt{\gamma_\circ^2-1}}\left[\frac{r_\circ}{r} 
- 1\right]\, . 
\end{split} 
\label{eqts_g_dopo_ad} 
\end{equation} 
\begin{equation} 
\begin{split} 
&\cos\vartheta=\frac{M_B  - m_i^\circ}{2r\sqrt{C}}\left( {r - r_\circ } 
\right) +\frac{m_i^\circ r_\circ }{8r\sqrt{C}}\left[ {\left( 
{\frac{r}{{r_\circ }}} \right)^4  - 1} \right] \\[6pt] 
&+\frac{{r_\circ \sqrt{C} }}{{12rm_i^\circ A^2 }} \ln \left\{ 
{\frac{{\left[ {A + \left(r/r_\circ\right)} \right]^3 \left(A^3  + 
1\right)}}{{\left[A^3  + \left( r/r_\circ \right)^3\right] \left( {A + 1} 
\right)^3}}} \right\} +\frac{ct_\circ}{r}-\frac{ct_a}{r} \\[6pt] & + 
\frac{r^\star}{r} +\frac{{r_\circ \sqrt{3C} }}{{6rm_i^\circ A^2 }} \left[ 
\arctan \frac{{2\left(r/r_\circ\right) - A}}{{A\sqrt{3} }} - \arctan 
\frac{{2 - A}}{{A\sqrt{3} }}\right]\, . 
\end{split} 
\label{eqts_g_dopo} 
\end{equation} 
The two EQTSs are represented at selected values of the arrival time $t_a$ in Fig.~\ref{eqts_comp}, where the illustrative case of GRB~991216 has been used as a prototype. The initial conditions at the beginning of the afterglow era are in this case given by $\gamma_\circ = 310.131$, $r_\circ = 1.943 \times 10^{14}$ cm, $t_\circ = 6.481 \times 10^{3}$ s, $r^\star = 2.354 \times 10^8$ cm (see Ruffini et al. \cite{lett1,lett2,rbcfx02_letter,Brasile}).

\subsection{The bolometric luminosity of the source} 

We assume that the internal energy due to kinetic collision is instantly radiated away and that the corresponding emission is isotropic. As in section \ref{eqaft}, let $\Delta \varepsilon$ be the internal energy density developed in the collision. In the comoving frame the energy per unit of volume and per solid angle is simply
\begin{equation} 
\left(\frac{dE}{dV d\Omega}\right)_{\circ}  =  \frac{\Delta \varepsilon}{4 
\pi} 
\label{dEo} 
\end{equation} 
due to the fact that the emission is isotropic in this frame. The total number of photons emitted is an invariant quantity independent of the frame used. Thus we can compute this quantity as seen by an observer in the comoving frame (which we denote with the subscript ``$\circ$'') and by an observer in the laboratory frame (which we denote with no subscripts). Doing this we find:
\begin{equation} 
\frac{dN_\gamma}{dt d \Omega d \Sigma}= \left(\frac{dN_\gamma}{dt d \Omega 
d \Sigma} \right)_{\circ} \Lambda^{-3} 
\cos \vartheta 
\, , 
\end{equation} 
where $\cos\vartheta$ comes from the projection of the elementary surface of the shell on the direction of propagation and $\Lambda = \gamma ( 1 - \beta \cos \vartheta )$ is the Doppler factor introduced in the two following differential transformation
\begin{equation} 
d \Omega_{\circ} = d \Omega \times \Lambda^{-2} 
\end{equation} 
for the solid angle transformation and 
\begin{equation} 
d t_{\circ} = d t \times \Lambda^{-1} 
\end{equation} 
for the time transformation. The integration in $d \Sigma$ is performed over the visible area of the ABM pulse at laboratory time $t$, namely with $0\le\vartheta\le\vartheta_{max}$ and $\vartheta_{max}$ defined in section \ref{eqts} (see Figs. \ref{openang}--\ref{opening}). An extra $\Lambda$ factor comes from the energy transformation:
\begin{equation} 
E_{\circ} = E \times \Lambda\, . 
\end{equation} 
See also Chiang and Dermer \cite{cd99}. Thus finally we obtain:
\begin{equation} 
\frac{dE}{dt d \Omega d \Sigma} = \left(\frac{dE}{dt d \Omega d \Sigma} 
\right)_{\circ} \Lambda^{-4} \cos \vartheta \, . 
\end{equation} 
Doing this we clearly identify  $  \left(\frac{dE}{dt d \Omega d \Sigma} \right)_{\circ} $ 
as the energy density in the comoving frame up to a factor $\frac{v}{4\pi}$ (see Eq.(\ref{dEo})). Then we have: 
\begin{equation} 
\frac{dE}{dt d \Omega } = \int_{shell} \frac{\Delta \varepsilon}{4 \pi} \; 
v \; \cos \vartheta \; \Lambda^{-4} \; d \Sigma\, , 
\label{fluxlab} 
\end{equation} 
where the integration in $d \Sigma$ is performed over the ABM pulse visible area at laboratory time $t$, namely with $0\le\vartheta\le\vartheta_{max}$ and $\vartheta_{max}$ defined in section \ref{eqts}. Eq.(\ref{fluxlab}) gives us the energy emitted toward the observer per unit solid angle and per unit laboratory time $t$ in the laboratory frame.

What we really need is the energy emitted per unit solid angle and per unit detector arrival time $t_a^d$, so we must use the complete relation between $t_a^d$ and $t$ given in Eq.(\ref{ta_g}). First we have to multiply the integrand in Eq.(\ref{fluxlab}) by the factor $\left(dt/dt_a^d\right)$ to transform the energy density generated per unit of laboratory time $t$ into the energy density generated per unit arrival time $t_a^d$. Then we have to integrate with respect to $d \Sigma$ over the {\em equitemporal surface} (EQTS, see section \ref{eqts}) of constant arrival time $t_a^d$ instead of the ABM pulse visible area at laboratory time $t$. The analog of Eq.(\ref{fluxlab}) for the source luminosity in detector arrival time is then:
\begin{equation} 
\frac{dE_\gamma}{dt_a^d d \Omega } = \int_{EQTS} \frac{\Delta 
\varepsilon}{4 \pi} \; v \; \cos \vartheta \; \Lambda^{-4} \; 
\frac{dt}{dt_a^d} d \Sigma\, . 
\label{fluxarr} 
\end{equation} 
It is important to note that, in the present case of GRB 991216, the Doppler factor $\Lambda^{-4}$ in Eq.(\ref{fluxarr}) enhances the apparent luminosity of the burst, as compared to the intrinsic luminosity, by a factor which at the peak of the afterglow is in the range between $10^{10}$ and $10^{12}$!
\begin{figure}
\centering
\includegraphics[width=\hsize,clip]{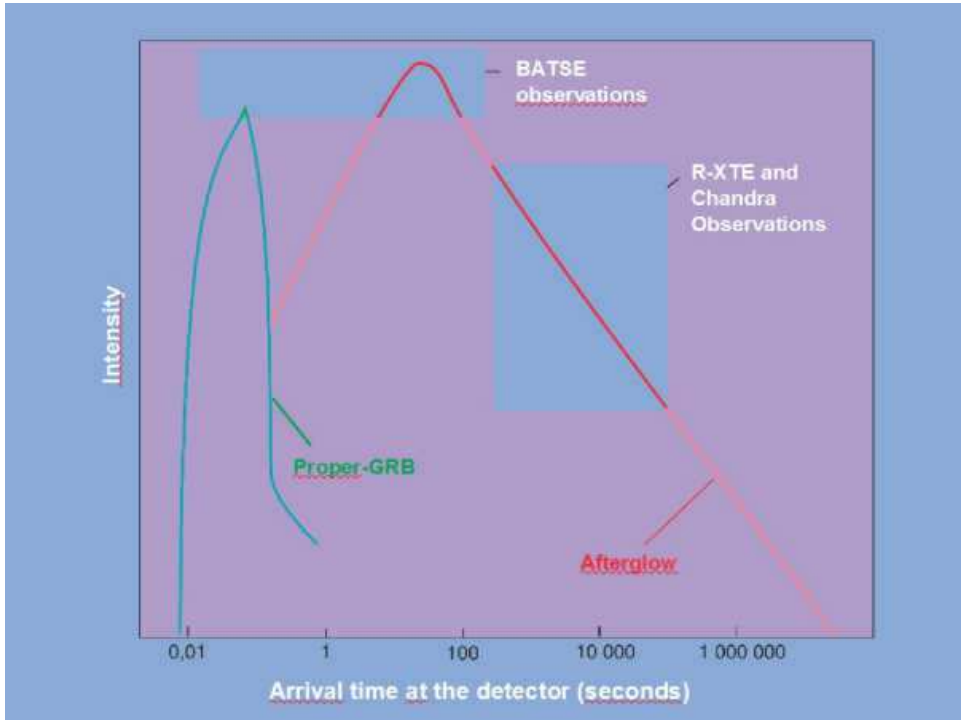}
\caption{Bolometric luminosity of P-GRB and afterglow as a function of the arrival time. Details in Ruffini et al. \cite{Brasile}. Reproduced and adapted from Ruffini et al. \cite{pls} with the kind permission of the publisher.}
\label{bolum}
\end{figure}

We are now able to reproduce in Fig. \ref{bolum} the general behavior of the luminosity starting from the P-GRB to the latest phases of the afterglow as a function of the arrival time. It is generally agreed that the GRB afterglow originates from an ultrarelativistic shell of baryons with an initial Lorentz factor $\gamma_\circ\sim 200$--$300$ with respect to the interstellar medium (see e.g. Ruffini et al. \cite{Brasile}, Bianco \& Ruffini \cite{EQTS_ApJL} and references therein). Using GRB 991216 as a prototype, in Ruffini et al. \cite{lett1,lett2} we have shown how from the time varying bolometric intensity of the afterglow it is possible to infer the average density $\left<n_{ism}\right>=1$ particle/cm$^3$ of the InterStellar Medium (ISM) in a region of approximately $10^{17}$ cm surrounding the black hole giving rise to the GRB phenomenon.

It was shown in Ruffini et al. \cite{rbcfx02_letter} that the theoretical interpretation of the intensity variations in the prompt phase in the afterglow implies the presence in the ISM of inhomogeneities of typical scale $10^{15}$ cm. Such inhomogeneities were there represented for simplicity as spherically symmetric over-dense regions with $\left<n_{ism}^{od}\right> \simeq 10^2\left<n_{ism}\right>$ separated by under-dense regions with $\left<n_{ism}^{ud}\right> \simeq 10^{-2}\left<n_{ism}\right>$ also of typical scale $\sim 10^{15}$ cm in order to keep $\left<n_{ism}\right>$ constant.

\begin{figure}
\centering
\includegraphics[width=\hsize,clip]{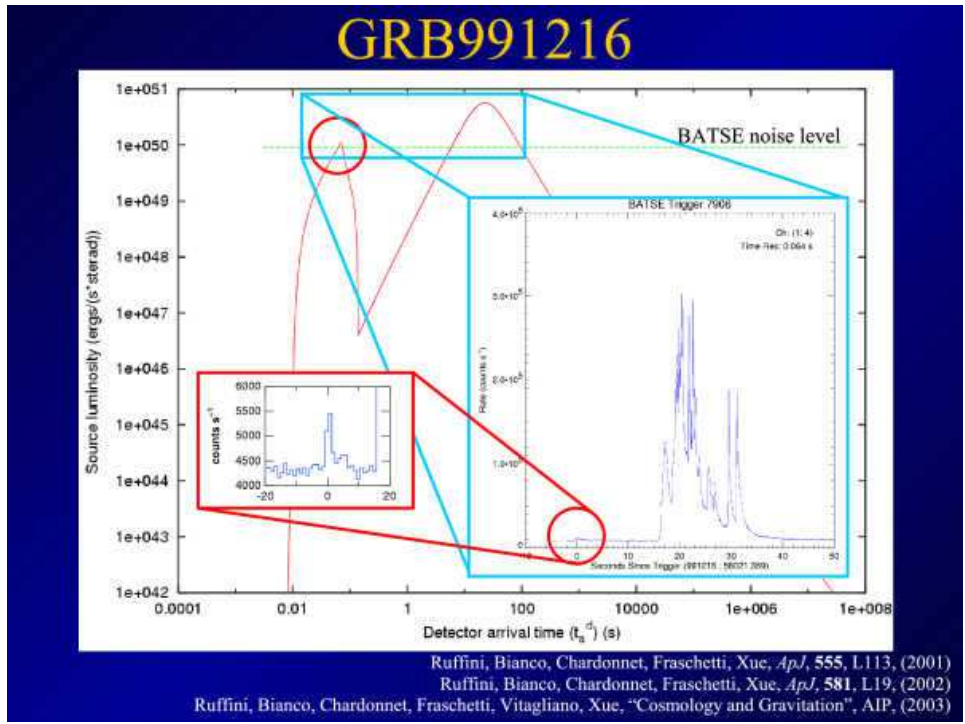}
\caption{The detailed features of GRB 991216 evidenced by our theoretical models are here reproduced. The P-GRB, the ``prompt radiation'' and what is generally called the afterglow. It is clear that the prompt radiation coincides with the extended afterglow peak emission (E-APE) and has been considered as a burst only as a consequence of the high noise threshold in the observations. Details in Ruffini et al. \cite{rbcfx02_letter,Brasile}.}
\label{grb991216}
\end{figure}

The summary of these general results are shown in Fig. \ref{grb991216}, where the P-GRB, the emission at the peak of the afterglow in relation to the ``prompt emission'' and the latest part of the afterglow are clearly identified for the source GRB 991216. Details in Ruffini et al. \cite{Brasile}.

\section{The theory of the luminosity in fixed energy bands and spectra of the afterglow} 

Having obtained a general agreement between the observed luminosity variability and our treatment of the bolometric luminosity, we have further developed the model in order to explain\\ 
a) the details  of the observed luminosity in fixed energy bands, which are the ones actually measured by the detectors on the satellites\\
b) the instantaneous as well as the average spectral distribution in the 
entire afterglow and\\ 
c) the observed hard to soft drift observed in GRB spectra. 

In order to do so we have developed (Ruffini et al. \cite{Spectr2}) a more detailed theory of the structure of the shock front giving rise to the afterglow. We have modeled the interaction between the ultrarelativistic shell of baryons and the ISM by a shock front with three well-defined layers (see e.g. secs. 85--89, 135 of Landau \& Lifshitz \cite{ll6}, ch. 2 and sec. 13--15 of Zel'dovich \& Rayzer \cite{zr66} and sec. IV, 11--13 of Sedov \cite{sedov}). From the back end to the leading edge of this shock front there is:\\ 
{\bf a)} A compressed high-temperature layer, of thickness $\Delta'$, in front of the relativistic baryonic shell, created by the accumulated material swept up in the ISM.\\ 
{\bf b)} A thin shock front, with a jump $\Delta T$ in the temperature which has been traditionally estimated in the comoving frame by the Rankine-Hugoniot adiabatic equations: 
\begin{equation} 
\Delta T \simeq (3/16) m_p\delta v^2/k \simeq 1.5\times 10^{11} 
\left[\delta v/(10^5 km s^{-1})\right]^2 K\,, 
\label{EqT} 
\end{equation} 
where $\delta v$ is the velocity jump, $m_p$ is the proton mass and $k$ is Boltzmann's constant. Of course such a treatment, valid for $\gamma \sim 1$, has to be modified (see below) in our novel treatment for the $\gamma \sim 200$ case relevant to GRBs.\\
{\bf c)} A pre-shock layer of ISM swept-up matter at much lower density and temperature, both of which change abruptly at the thin shock front behind it.

At larger distances ahead of the expanding fireball the ISM is at still smaller densities. The upper limit to the temperature jump at the thin shock front, given in Eq.(\ref{EqT}), is due to the transformation of kinetic energy to thermal energy, since the particle mean free path is assumed to be less then the thickness of the layer (a). The thermal emission of the observed X- and gamma ray radiation, which as seen from the observations reveals a high level of stability, is emitted in the above region (a) due to the sharp temperature gradient at the thin shock front described in the above region (b).

The optical and radio emission comes in our model from the extended region (c). The description of such a region, unlike the sharp and well-defined temperature gradient occurring in region (b), requires magnetohydrodynamic simulations of the evolution of the electron energy distribution of the synchrotron emission. Such analysis has been performed using 3-D Eulerian MHD codes for the particle acceleration models to produce the energy spectrum of cosmic rays at supernova envelope fronts (see e.g. McKee and Cowie \cite{mc75}, Tenorio-Tagle et al. \cite{tt91}, Stone and Norman \cite{sn92}, Jun \& Jones \cite{j99}). Other challenges are the magnetic field and the instabilities. We mention two key phenomena: first, the importance of the development of Kelvin-Helmholtz and Rayleigh-Taylor instabilities ahead of the thin shock front. The second is the dual effect that the shock front has on the ISM initial magnetic field, first through the compression of the swept-up matter containing the field and secondly the amplification of the radial magnetic field component due to the Rayleigh-Taylor instability. Simulations of both effects (see e.g. Jun and Jones \cite{j99} and references therein), modeling the synchrotron radio emission for an expanding supernova shell at various initial magnetic field and ISM parameter values, shows for example that the presence of an initial tangential magnetic field component may essentially affect the resulting magnetic field configuration and hence the outgoing radio flux and spectrum. Among the additional effects to be taken into account are the initial inhomogeneity of the ISM and the contribution of magnetohydrodynamic turbulence.

In our approach we focus uniquely on the X- and gamma ray radiation, which appears to be conceptually much simpler than the optical and radio emission. It is perfectly predictable by a set of constitutive equations (see next section), which leads to directly verifiable and very stable features in the spectral distribution of the observed GRB afterglows. In line with the observations of GRB 991216 and other GRB sources, we assume in the following that the X- and gamma ray luminosity represents approximately 90\% of the energy flux of the afterglow, while the optical and radio emission represents only the remaining 10\%. 

This approach differs significantly from the other ones in the current literature, where attempts are made to explain at once all the multi-wavelength emission in the radio, optical, X and gamma ray as coming from a common origin which is linked to boosted synchrotron emission. Such an approach has been shown to have a variety of difficulties (Ghirlanda et al. \cite{gcg02}, Preece et al. \cite{pa98}) and cannot anyway have the instantaneous variability needed to explain the structure in the ``prompt radiation'' in an external shock scenario, which is indeed confirmed by our model.

\subsection{The equations determining the luminosity in fixed energy bands} 

Here the fundamental new assumption is adopted (see also Ruffini et al. \cite{Spectr1}) that the X- and gamma ray radiation during the entire afterglow phase has a thermal spectrum in the co-moving frame. The temperature is then given by:
\begin{equation} 
T_s=\left[\Delta E_{\rm int}/\left(4\pi r^2 \Delta \tau \sigma 
\mathcal{R}\right)\right]^{1/4}\, , 
\label{TdiR} 
\end{equation} 
where $\Delta E_{\rm int}$ is the internal energy developed in the collision with the ISM in a time interval $\Delta \tau$ in the co-moving frame, $\sigma$ is the Stefan-Boltzmann constant and
\begin{equation} 
\mathcal{R}=A_{eff}/A\, , 
\label{Rdef} 
\end{equation} 
is the ratio between the ``effective emitting area'' of the afterglow and the surface area of radius $r$. In GRB 991216 such a factor is observed to be decreasing during the afterglow between: $3.01\times 10^{-8} \ge \mathcal{R} \ge 5.01 \times 10^{-12}$ (Ruffini et al. \cite{Spectr1}). 

\begin{figure}
\includegraphics[width=\hsize,clip]{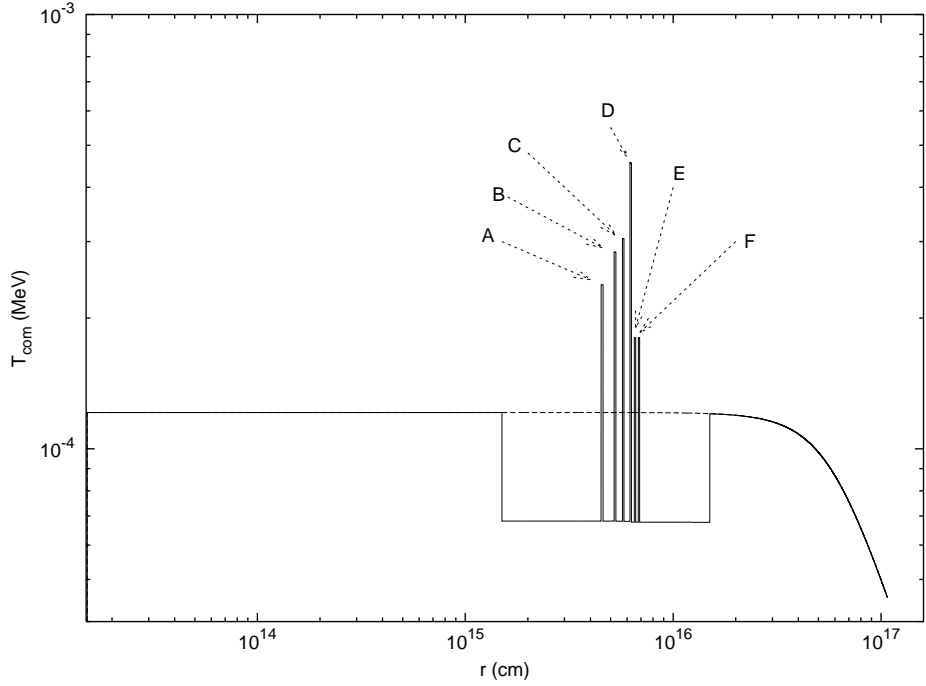}
\caption{The temperature in the comoving frame of the shock front corresponding to the density distribution with the six spikes A,B,C,D,E,F presented in Ruffini et al.$^5$. The dashed line corresponds to an homogeneous distribution with $n_{ism}=1$. Details in Ruffini et al. \cite{Spectr2}.}
\label{tcom_fig}
\end{figure}

The temperature in the comoving frame corresponding to the density distribution described in Ruffini et al. \cite{rbcfx02_letter} is shown in Fig. \ref{tcom_fig}. 

We are now ready to evaluate the source luminosity in a given energy band. The source luminosity at a detector arrival time $t_a^d$, per unit solid angle $d\Omega$ and in the energy band $\left[\nu_1,\nu_2\right]$ is given by (see Ruffini et al. \cite{Brasile,Spectr1}): 
\begin{equation} 
\frac{dE_\gamma^{\left[\nu_1,\nu_2\right]}}{dt_a^d d \Omega } = 
\int_{EQTS} \frac{\Delta \varepsilon}{4 \pi} \; v \; \cos \vartheta \; 
\Lambda^{-4} \; \frac{dt}{dt_a^d} W\left(\nu_1,\nu_2,T_{arr}\right) d 
\Sigma\, , 
\label{fluxarrnu} 
\end{equation} 
where $\Delta \varepsilon=\Delta E_{int}/V$ is the energy density released in the interaction of the ABM pulse with the ISM inhomogeneities measured in the comoving frame, $\Lambda=\gamma(1-(v/c)\cos\vartheta)$ is the Doppler factor, $W\left(\nu_1,\nu_2,T_{arr}\right)$ is an ``effective weight'' required to evaluate only the contributions in the energy band $\left[\nu_1,\nu_2\right]$, $d\Sigma$ is the surface element of the EQTS at detector arrival time $t_a^d$ on which the integration is performed (see also Ruffini et al. \cite{rbcfx02_letter}) and $T_{arr}$ is the observed temperature of the radiation emitted from $d\Sigma$: 
\begin{equation} 
T_{arr}=T_s/\left[\gamma 
\left(1-(v/c)\cos\vartheta\right)\left(1+z\right)\right]\, . 
\label{Tarr} 
\end{equation} 

The ``effective weight'' $W\left(\nu_1,\nu_2,T_{arr}\right)$ is given by the ratio of the integral over the given energy band of a Planckian distribution at a temperature $T_{arr}$ to the total integral $aT_{arr}^4$: 
\begin{equation} 
W\left(\nu_1,\nu_2,T_{arr}\right)=\frac{1}{aT_{arr}^4}\int_{\nu_1}^{\nu_2}\rho\left(T_{arr},\nu\right)d\left(\frac{h\nu}{c}\right)^3\, , 
\label{effweig} 
\end{equation} 
where $\rho\left(T_{arr},\nu\right)$ is the Planckian distribution at temperature $T_{arr}$: 
\begin{equation} 
\rho\left(T_{arr},\nu\right)=\left(2/h^3\right)h\nu/\left(e^{h\nu/\left(kT_{arr}\right)}-1\right) 
\label{rhodef} 
\end{equation} 

\section{On the time integrated spectra and the hard-to-soft spectral transition} 

\begin{figure}
\centering
\includegraphics[width=\hsize,clip]{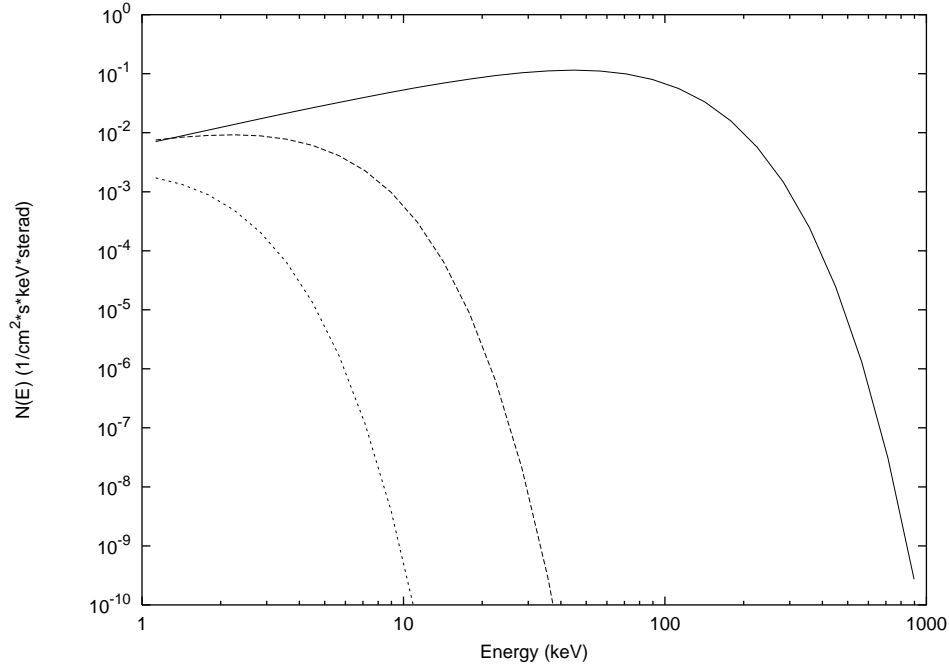} 
\caption{The instantaneous spectra of the radiation observed in GRB~991216 at three different EQTS respectively, from top to bottom, for $t_a^d=10$ s, $t_a^d=10^4$ s and $t_a^d=1.45\times10^5$ s. These diagrams have been computed assuming a constant $\left<n_{ism}\right>\simeq 1$ particle/cm$^3$ and clearly explains the often quoted hard-to-soft spectral evolution in GRBs. Details in Ruffini et al. \cite{Spectr1}.}
\label{spectrum} 
\end{figure}

We turn now to the much debated issue of the origin of the observed hard-to-soft spectral transition during the GRB observations (see e.g. Frontera et al. \cite{fa00}, Ghirlanda et al. \cite{gcg02}, Piran \cite{p99}, Piro et al. \cite{p99b}). We consider the instantaneous spectral distribution of the observed radiation for three different EQTSs:
\begin{itemize} 
\item $t_a^d=10$ s, in the early radiation phase near the peak of the luminosity, 
\item $t_a^d=1.45\times 10^5$ s, in the last observation of the afterglow by the Chandra satellite, and 
\item $t_a^d=10^4$ s, chosen in between the other two (see Fig.~\ref{spectrum}). 
\end{itemize} 
The observed hard-to-soft spectral transition is then explained and traced back to: 
\begin{enumerate} 
\item a time decreasing temperature of the thermal spectrum measured in the comoving frame, 
\item the GRB equations of motion, 
\item the corresponding infinite set of relativistic transformations. 
\end{enumerate} 
A clear signature of our model is the existence of a common low-energy behavior of the instantaneous spectrum represented by a power-law with index $\alpha = +0.9$. This prediction will be possibly verified in future observations.

\begin{figure}
\centering
\includegraphics[width=\hsize,clip]{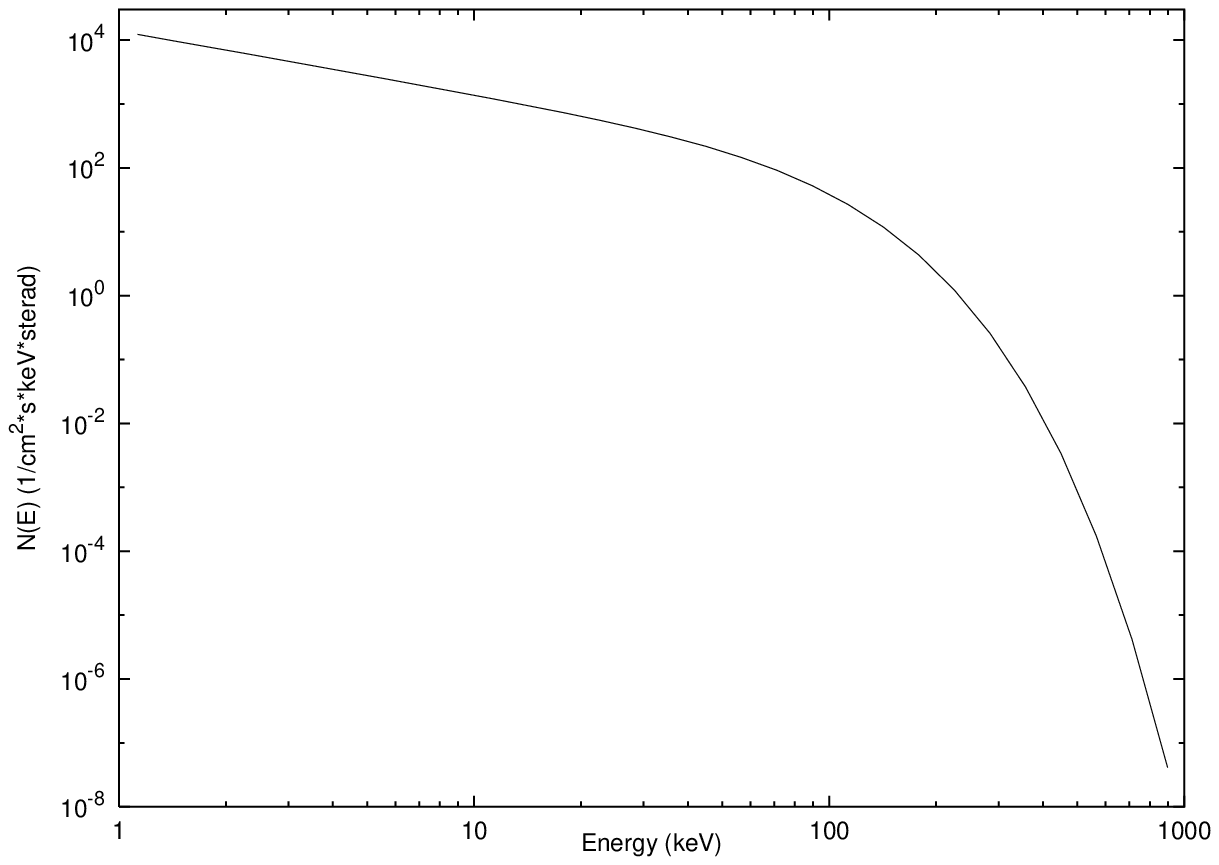} 
\caption{The time-integrated spectrum of the radiation observed in GRB~991216. The low energy part of the curve below $10$ keV is fit by a power-law with index $\alpha = -1.05$ and the high energy part above $500$ keV is fit by a power-law with an index $\beta < -16$. Details in Ruffini et al. \cite{Spectr1}.}
\label{spectband} 
\end{figure}

Starting from these instantaneous values, we integrate the spectra in arrival time obtaining what is usually fit in the literature by the ``Band relation'' (Band et al. \cite{b93}). Indeed we find for our integrated spectra a low energy spectral index $\alpha=-1.05$ and an high energy spectral index $\beta < -16$ when interpreted within the framework of a Band relation (see Fig.~\ref{spectband}). This theoretical result can be submitted to a direct confrontation with the observations of GRB 991216 and, most importantly, the entire theoretical framework which we have developed can now be applied to any GRB source. The theoretical predictions on the luminosity in fixed energy bands so obtained can be then straightforwardly confronted with the observational data.

\section{The three paradigms for the interpretation of GRBs} 

Having outlined the main features of our model and shown its application to GRB 991216 used as a prototype, before addressing the two new sources which are going to be the focus of this presentation, we recall the three paradigms for the interpretation of GRBs we had previously introduced.

The first paradigm, the relative space-time transformation (RSTT) paradigm (Ruffini et al. \cite{lett1}) emphasizes the importance of a global analysis of the GRB phenomenon encompassing both the optically thick and the afterglow phases. Since all the data are received in the detector arrival time it is essential to know the equations of motion of all relativistic phases with $\gamma > 1$ of the GRB sources in order to reconstruct the time coordinate in the laboratory frame, see Eq.(\ref{tadef}). Contrary to other phenomena in nonrelativistic physics or astrophysics, where every phase can be examined separately from the others, in the case of GRBs all the phases are inter-related by their signals received in arrival time $t_a^d$. There is the need, in order to describe the physics of the source, to derive the laboratory time $t$ as a function of the arrival time $t_a^d$ along the entire past worldline of the source using Eq.(\ref{taddef}).

\begin{figure}
\centering
\includegraphics[width=\hsize,clip]{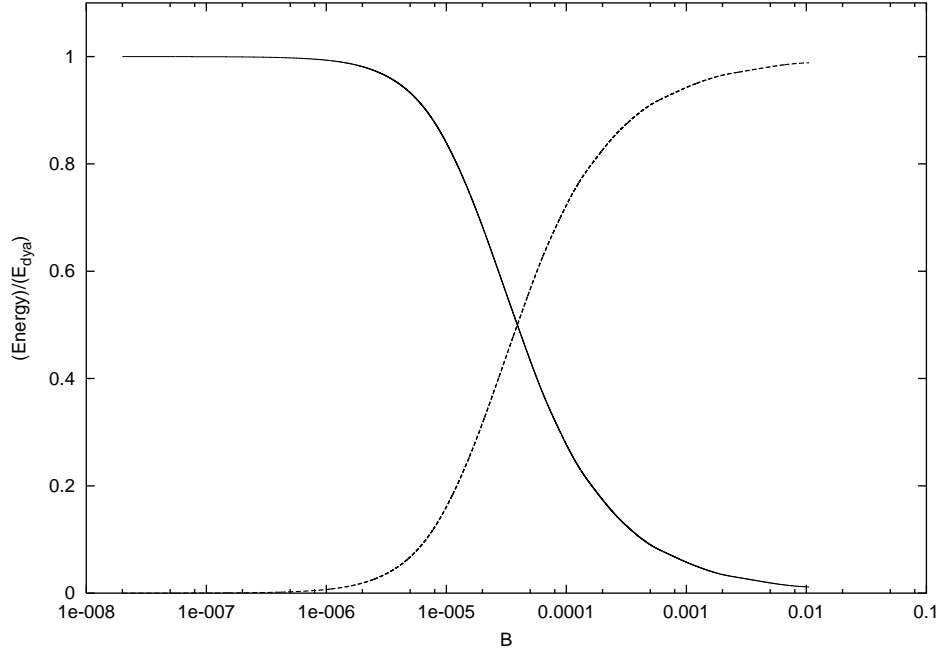}
\caption{The energy radiated in the P-GRB (the solid line) and in the afterglow (the dashed line), in units of the total energy of the dyadosphere ($E_{dya}$), are plotted as functions of the $B$ parameter.}
\label{bcross}
\end{figure}

The second paradigm, the interpretation of the burst structure (IBS) paradigm (Ruffini et al. \cite{lett2}) covers three fundamental issues:\\ 
a) the existence, in the general GRB, of two different components: the P-GRB and the afterglow related by precise equations determining their relative amplitude and temporal sequence (see Ruffini et al. \cite{Brasile});\\ 
b) what in the literature has been addressed as the ``prompt emission'' and considered as a burst, in our model is not a burst at all --- instead it is just the emission from the peak of the afterglow (see Fig. \ref{grb991216});\\
c) the crucial role of the parameter $B$ in determining the relative amplitude of the P-GRB to the afterglow and discriminating between the short and the long bursts (see Fig. \ref{bcross}). Both short and long bursts arise from the same physical phenomena: the dyadosphere. The absence of baryonic matter in the remnant leads to the short bursts and no afterglow. The presence of baryonic matter with $B < 10^{-2}$ leads to the afterglow and consequently to its peak emission which gives origin to the so-called long bursts.

The third paradigm, the GRB-Supernova Time Sequence (GSTS) paradigm (Ruffini et al. \cite{lett3}), deals with the relation of the GRB and the associated supernova process, and acquires a special meaning in relation to the sources GRB 980425 and GRB 030329 as we will show in the following.

We now shortly illustrate some consequences of these three paradigms.

\subsection{Long bursts are E-APEs}\label{eape}

The order of magnitude estimate usually quoted for the the characteristic time scale to be expected for a burst emitted by a GRB at the moment of transparency at the end of the expansion of the optically thick phase is given by $\tau \sim GM/c^3$, which for a $10M_\odot$ black hole will give $\sim 10^{-3}$ s. There are reasons today not to take seriously such an order of magnitude estimate (see e.g. Ruffini et al. \cite{rfvx05}). In any case this time is much shorter then the ones typically observed in ``prompt radiation'' of the long bursts, from a few seconds all the way to $10^2$ s. In the current literature (see e.g. Piran \cite{p99} and references therein), in order to explain the ``prompt radiation'' and overcome the above difficulty it has been generally assumed that its origin should be related to a prolonged ``inner engine'' activity preceding the afterglow which is not well identified.

To us this explanation has always appeared logically inconsistent since there remain to be explained not one but two very different mechanisms, independent of each other, of similar and extremely large energetics. This approach has generated an additional very negative result: it has distracted everybody working in the field from the earlier very interesting work on the optically thick phase of GRBs.

The way out of this dichotomy in our model is drastically different: 1) indeed the optically thick phase exists, is crucial to the GRB phenomenon and terminates with a burst: the P-GRB; 2) the ``prompt radiation'' follows the P-GRB; 3) the ``prompt radiation'' is not a burst: it is actually the temporally extended peak emission of the afterglow (E-APE). The observed structures of the prompt radiation can all be traced back to inhomogeneities in the interstellar medium (see Fig. \ref{grb991216} and Ruffini et al. \cite{rbcfx02_letter}).

\subsection{Short bursts are P-GRBs} 

The fundamental diagram determining the relative intensity of the P-GRB and the afterglow as a function of the dimensionless parameter $B$ has been shown in Fig. \ref{bcross}. The underlying machine generating the short and the long GRBs is identical: in both cases is the dyadosphere. The main difference relates to the amount of baryonic matter engulfed by the electron-positron plasma in their optically thick phase prior to transparency. In the limit of small $B < 10^{-5}$ the intensity of the P-GRB is larger and dominates the afterglow. This corresponds to the short bursts. For $10^{-5} < B < 10^{-2}$ the  afterglow dominates the GRBs and we have the so-called ``long bursts''. For $B > 10^{-2}$ we may observe a third class of ``bursts'', eventually related to a turbulent process occurring prior to transparency (Ruffini et al. \cite{rswx00}). This third family should be characterized  by  smaller values of the Lorentz gamma factors than in the case of the short or long bursts.

\subsection{The trigger of multiple gravitational collapses}

The relation between the GRBs and the supernovae is one of the most complex aspects to be addressed by our model, which needs the understanding of new fields of general relativistic physics in relation to yet unexplored many-body solutions in a substantially new astrophysical scenario.

As we will show in the two systems GRB980425/SN1998bw and GRB030329/SN2003dh which we are going to discuss next, there is in each one the possibility of an astrophysical ``triptych''\footnote{A picture or carving in three panels side by side; {\em esp}: an altarpiece with a central panel and two flanking panels half its size that fold over it [Webster's New collegiate dictionary, G. \& C. Merriam Co. (Springfield, Massachussets, U.S.A., 1977)]} formed by:\\
1) the formation of the black hole and the emission of the GRB,\\ 
2) the gravitational collapse of an evolved companion star, leading to a supernova,\\ 
3) a clearly identified URCA source whose nature appears to be of the greatest interest.

This new astrophysical scenario presents new challenges:\\
a) The identification of the physical reasons of the instability leading to the gravitational collapse of a $\sim 10M_\odot$ star, giving origin to the black hole. Such an implosion must occur radially with negligible mass of the remnant ($B < 10^{-2}$).\\
b) The identification of the physical reasons for the instability leading to the gravitational collapse of an evolved companion star, giving origin to the supernova.\\
c) The theoretical issues related to the URCA sources, which range today in many possible directions: from the physics of black holes, to the physical processes occurring in the expanding supernova remnants, and finally to the very exciting possibility that we are observing for the first time a newly born neutron star. The main effort in the next sections is to show that the detailed understanding we have reached for the GRB phenomenon and its afterglow allows us to state, convincingly, that the URCA source, contrary to what established in the current literature, is not part of the GRB nor of its afterglow.

We will draw in the conclusions some considerations on the possible nature of the URCA sources.

\section{Applications} 

We illustrate the application of our GRB model to two different systems, which are quite different in the energetics but are both related to supernovae: GRB 980425 and GRB 030329. We will let the gradual theoretical understanding of the system to unveil the underlying astrophysical scenario.

\subsection{GRB 980425 / SN 1998bw}

Approaches in the current literature have always attempted to explain both the supernova and the GRB as two aspects of a single phenomenon assuming that the GRB takes his origin from a specially strong and yet unobserved supernova process: a hypernova (see Paczy\'nski \cite{p98}, Kulkarni \cite{ka98}, Iwamoto \cite{ia98}).

We have taken a very different approach, following Cicero's classic aphorism ``divide et impera'', which was adopted as the motto of the Roman empire: ``divide and conquer''. In this specific case of GRBs, which are indeed a very complex system, we plan to divide and identify the truly independent physical constituents and conquer the understanding of the underlying astrophysical process. As we will see, this approach will lead to an unexpected and much richer scenario.

\begin{figure}
\centering
\includegraphics[width=\hsize]{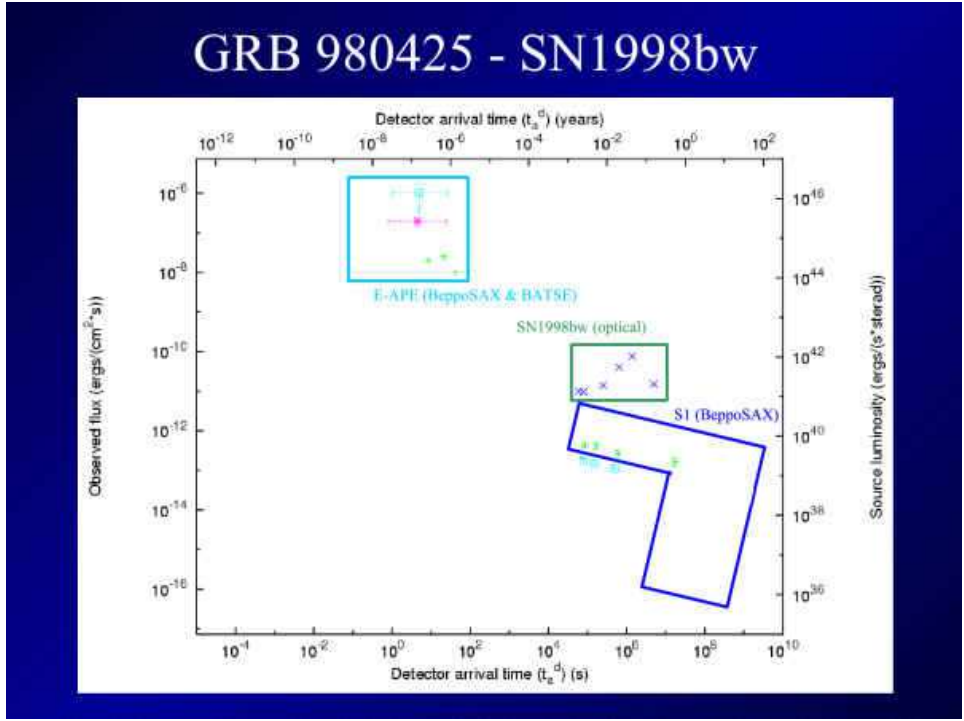}
\caption{The ``divide et impera'' concept applied to the system GRB 980425 / SN 1998bw. Four different components are identified: the GRB 980425, the SN 1998bw, and the two sources S1 and S2.}
\label{d14}
\end{figure}

In addition to the source GRB 980425 and the supernova SN1998bw, two X-ray sources have been found by BeppoSAX in the error box for the location of GRB 980425: a source {\em S1} and a source {\em S2} (Pian et al. \cite{pian00}), which have been traditionally interpreted either as a background source or as a part of the GRB afterglow. See Fig. \ref{d14}. Our approach has been: to  first comprehend the entire afterglow of GRB 980425 within our theory. This allows the computation of the luminosity in given energy bands, the spectra, the Lorentz gamma factors, and more generally of all the dynamical aspects of the source. Having characterized the features of GRB 980425, we can gradually approach the remaining part of the scenario, disentangling the GRB observations from those of the supernova and then disentangling both the GRB and the supernova observations from those of the sources {\em S1} and {\em S2}. This leads to a natural identification of distinct events and to their autonomous astrophysical characterization.

\begin{figure}
\centering 
\includegraphics[width=\hsize]{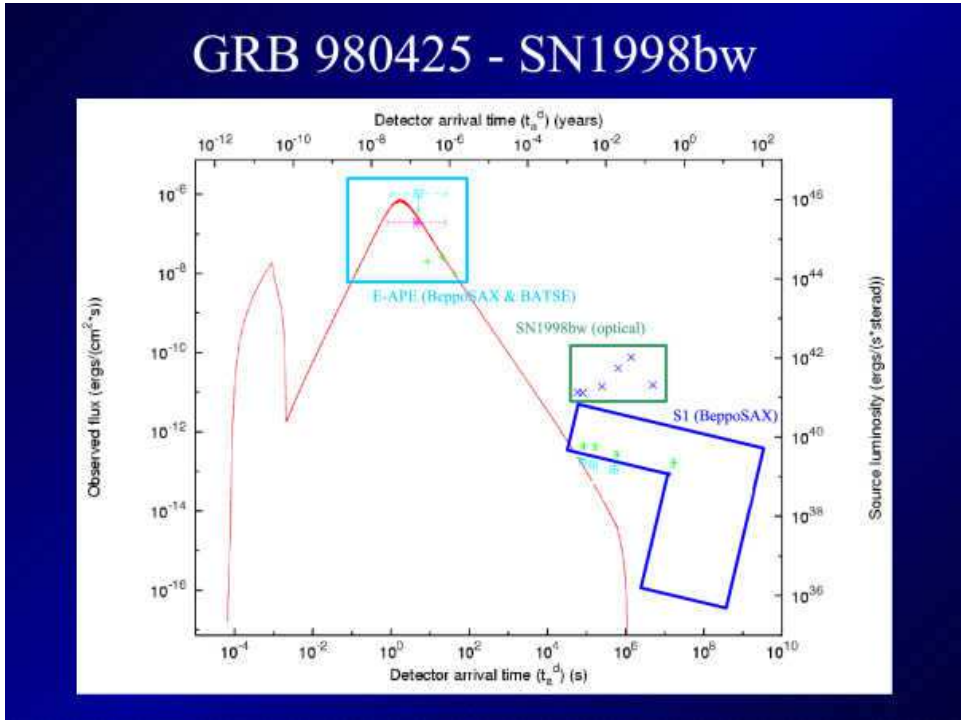}
\caption{The bolometric luminosity as a function of the arrival time. The peak of the P-GRB is just below the observational noise level.}
\label{d15} 
\end{figure}

\begin{figure}
\centering 
\includegraphics[width=\hsize]{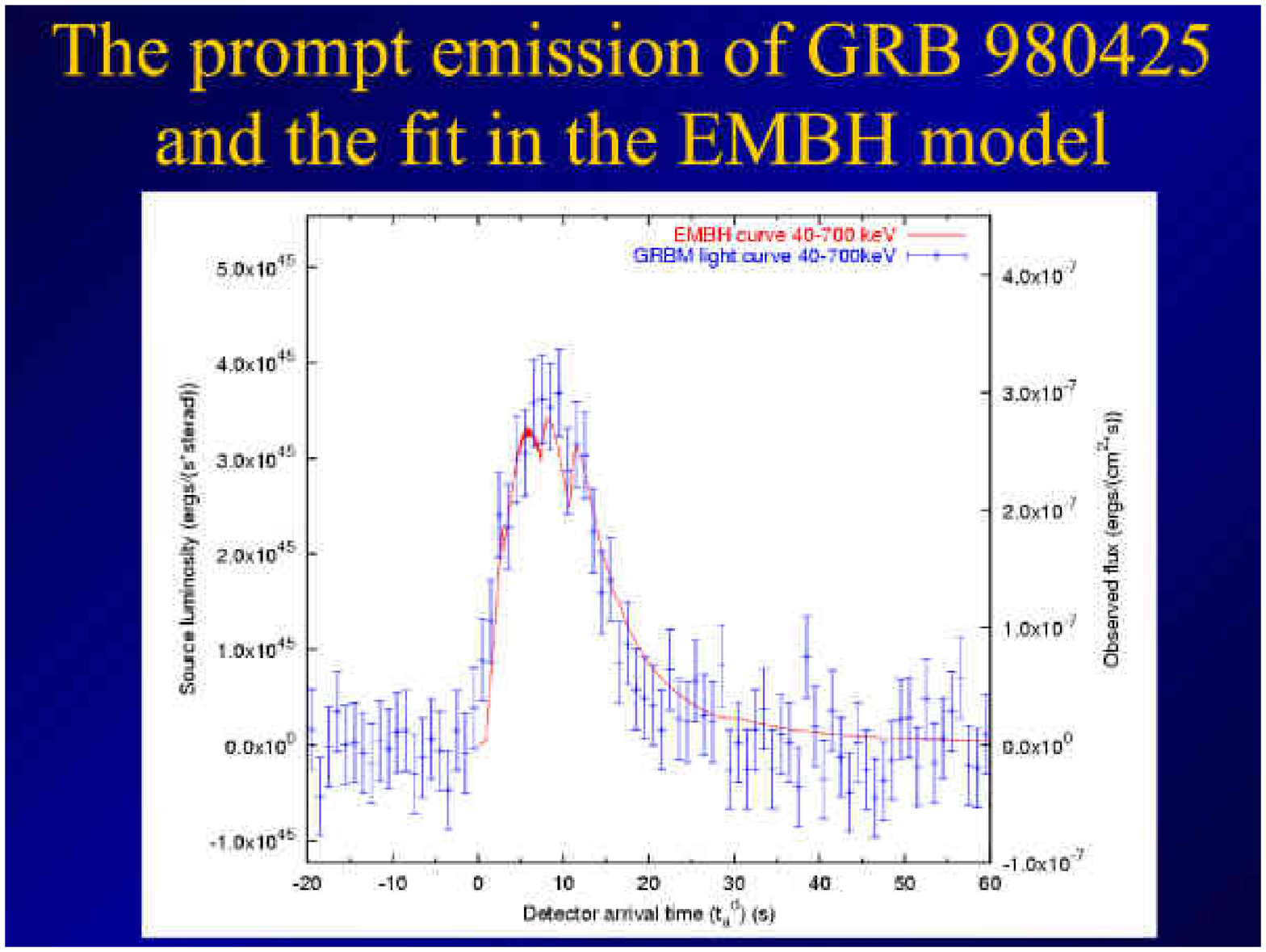}
\caption{The observation by BeppoSAX of the peak of the afterglow in the $40$--$700$ keV energy band is fitted by our model.}
\label{d17} 
\end{figure}

Our best fit for GRB 980425 corresponds to $E_{dya}=1.1\times 10^{48}\, {\rm ergs}$, $B=7\times 10^{-3}$ and the ISM average density is found to be $\left<n_{ism}\right>=0.02~{\rm particle}/{\rm cm}^3$. The plasma temperature and the total number of pairs in the dyadosphere are respectively $T=1.028\, {\rm MeV}$ and $N_{e^\pm}=5.3274\times10^{53}$. The light curve of the GRB is shown in Figs. \ref{d15}--\ref{d17}. The P-GRB is under the threshold and in the case of this source is not observable (see Ruffini et al. \cite{cospar02}, Fraschetti et al. \cite{f03mg10}).

\begin{figure}
\centering 
\includegraphics[width=\hsize]{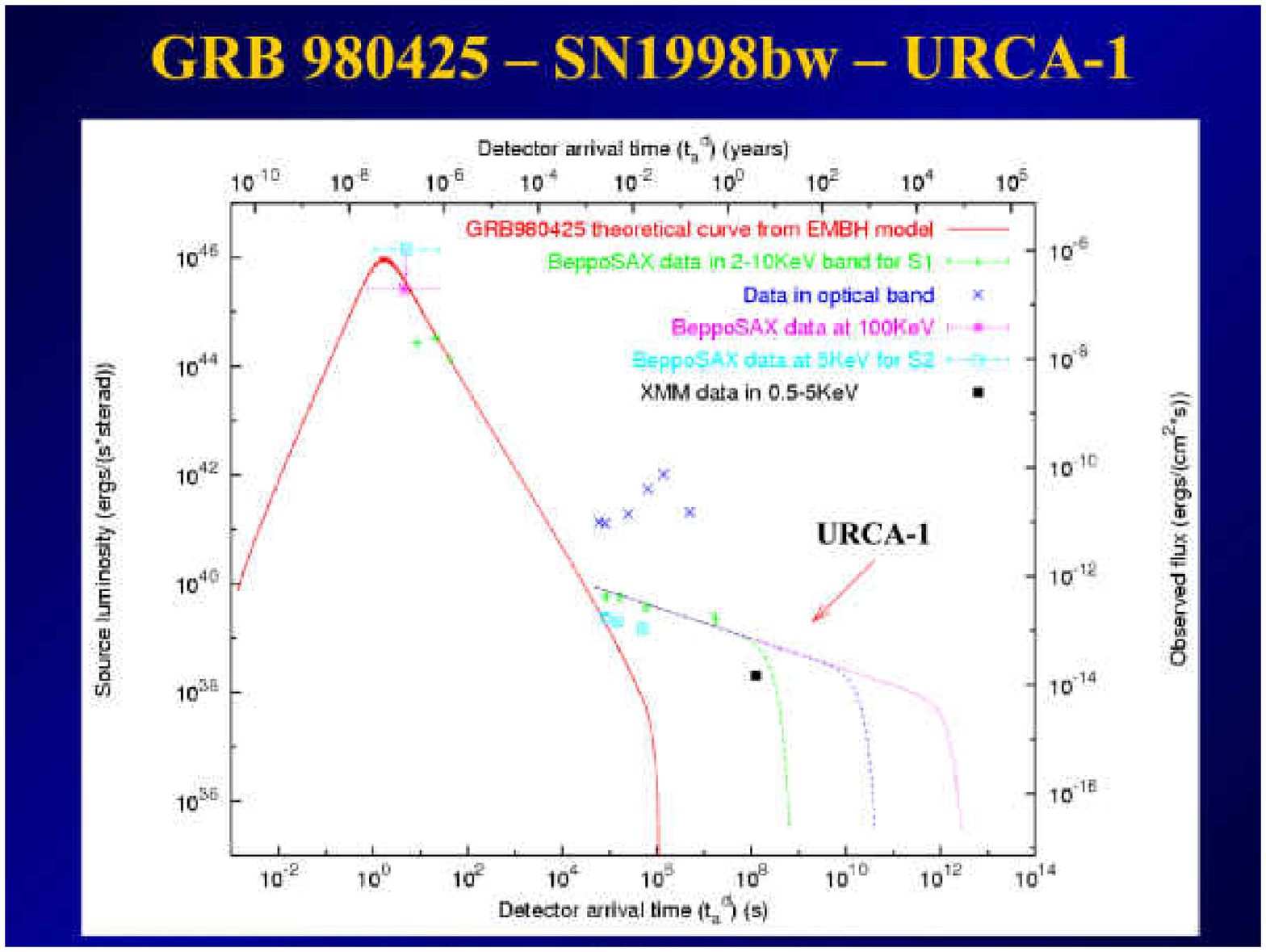} 
\caption{The bolometric light curves is reported as well as the BeppoSAX MECS observations in the $2$--$10$ keV band of S1 and S2 (Pian et al. \cite{pian00}) and the optical data of SN 1998bw (Iwamoto \cite{i99}). Here we also report some very qualitative curves to be expected for the URCA-1 luminosity. Details in Ruffini et al. \cite{cospar02}.}
\label{d19} 
\end{figure}

The characteristic parameter $\mathcal{R}$, defining the filamentary structure of the ISM, monotonically decreases from $4.81\times 10^{-10}$ to $2.65\times 10^{-12}$). The results are given in Fig. \ref{d19} where the bolometric luminosity is represented together with the optical data of SN1998bw, the source {\em S1} and the source {\em S2}. It is then clear that GRB 980425 is separated both from the supernova data and from the sources {\em S1} and {\em S2}. 

While the occurrence of the supernova in relation to the GRB has already been discussed within the GRB-Supernova Time Sequence (GSTS) paradigm (Ruffini et al. \cite{lett3}), we like to address here a different fundamental issue: the nature of the source {\em S1} which we have named, in celebration of the work of Gamow and Shoenberg, URCA-1. It is clear, from the theoretical predictions of the afterglow luminosity, that the URCA-1 cannot be part of the afterglow (see Figs. \ref{d15}, \ref{d19}). There are three different possibilities for the explanation of such source:\\
1) Its possible relation to the black hole formed during the process of gravitational collapse leading to the GRB emission.\\
2) Its possible relation to emission originating in the early phases of the expansion of the supernova remnant.\\
3) The very exciting possibility that for the first time we are observing a newly born neutron star out of the supernova phenomenon.

While some general considerations will be discussed in the conclusions, we would like to stress here the paramount importance of following the further time history of URCA-1 and of the source {\em S2}. If, as we propose, {\em S2} is a background source, its flux should be practically constant in time and this source has nothing to do with the GRB 980425 / SN1998bw system. The drastic behavior of the URCA-1 luminosity reported in the talk by Elena Pian in this meeting, showing the latest URCA-1 observations by the XMM and Chandra satellites, is crucial for the understanding of the nature of this source. Some very qualitative luminosity curves are sketched in Fig. \ref{d19}, illustrating the possible time evolution of URCA-1. They are still very undetermined today due to a lack of attention to these observational data and, consequently, to the lack of a detailed theoretical model of the phenomenon. We therefore propose to have a dedicated attention to the astrophysical ``triptych'' GRB 980425 / SN 1998bw / URCA-1.

\subsection{GRB 030329 / SN 2003dh}

We have adopted for our modeling of GRB030329 a spherically symmetric distribution for the source and, as initial conditions at $t = 10^{-21}$ s, an $e^+$-$e^-$-photon neutral plasma lying between the radii $r_1 = 2.9\times 10^6$ cm and $r_2 = 9.0\times 10^7$ cm. The temperature of such a plasma is 2.1 MeV, the total energy $E_{tot} = 2.1\times 10^{52}$ erg and the total number of pairs $N_{e^+e^-} = 1.1\times 10^{57}$. The baryonic matter component $M_B$ is the second free parameter of our theory: $B = 4.8 \times 10^{-3}$. At the emission of the P-GRB, the Lorentz gamma factor is $\gamma_\circ = 183.6$ and the radial coordinate is $r_\circ = 5.3 \times 10^{13}$ cm. The ISM average density is best fit by $<n_{ism}> = 1$ particle/cm$^3$. The third free parameter of our theory is given by $1.1\times 10^{-7} < \mathcal{R} < 5.0 \times 10^{-11}$.

\begin{figure}
\centering
\includegraphics[width=\hsize]{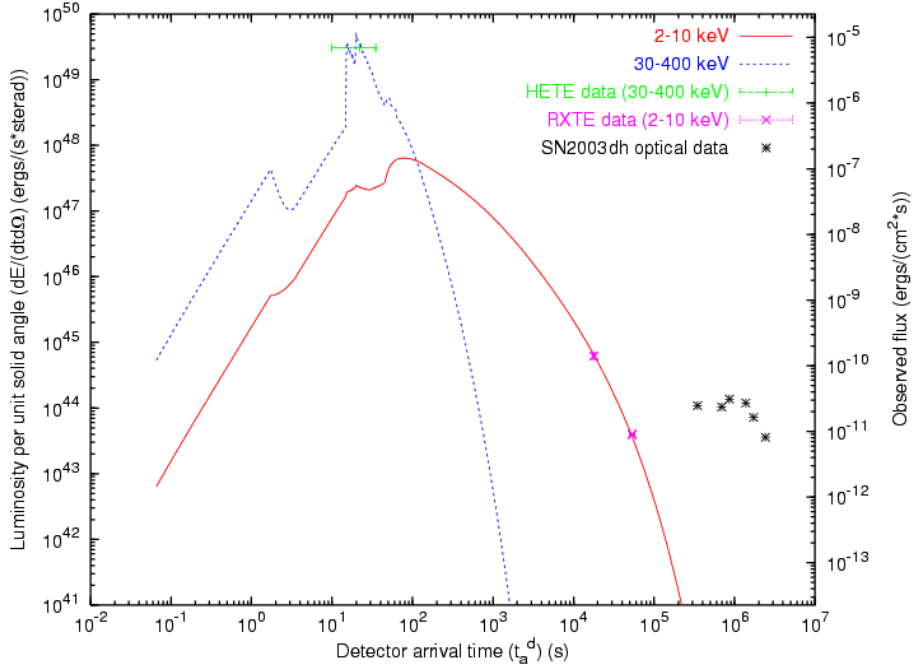} 
\caption{The luminosity in the $2$--$10$ keV and in the $30$--$400$ keV energy bands predicted by our model are fitted to the data of R-XTE (GCN Circ. 1996 \cite{gcn1996}) and HETE-2 (GCN Circ. 1997 \cite{gcn1997}) respectively. The SN 2003dh optical luminosity is given by the crosses (Hjorth et al. \cite{ha03}). Details in Bernardini et al. \cite{b03mg10}.}
\label{030329global} 
\end{figure}

\begin{figure}
\centering
\includegraphics[width=\hsize]{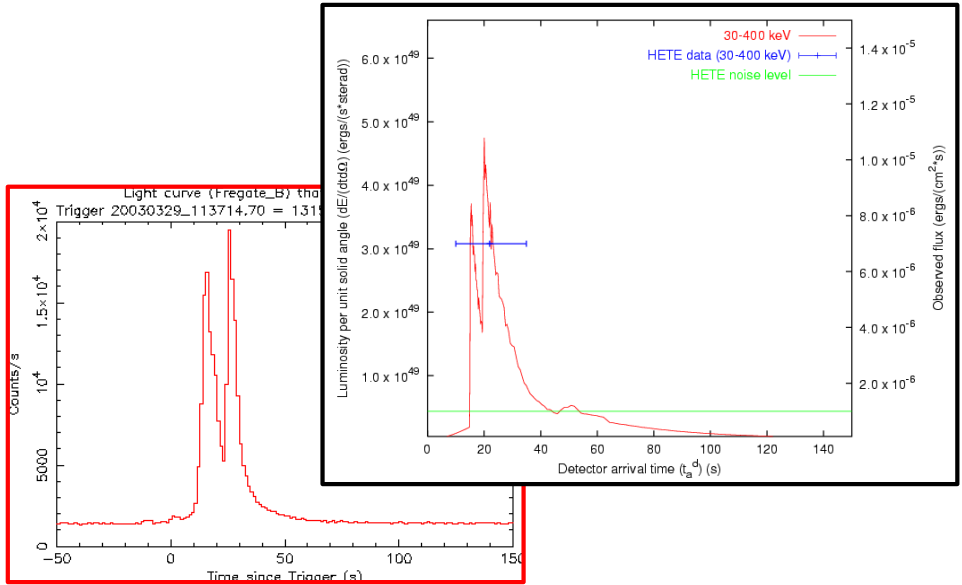} 
\caption{The details of the theoretical fit of the prompt radiation of GRB 030329 have been reproduced by the filamentary structure in the ISM in our model. Details in Bernardini et al. \cite{b03mg10}.} 
\label{030329sptp} 
\end{figure}

\begin{figure}
\centering
\includegraphics[width=\hsize]{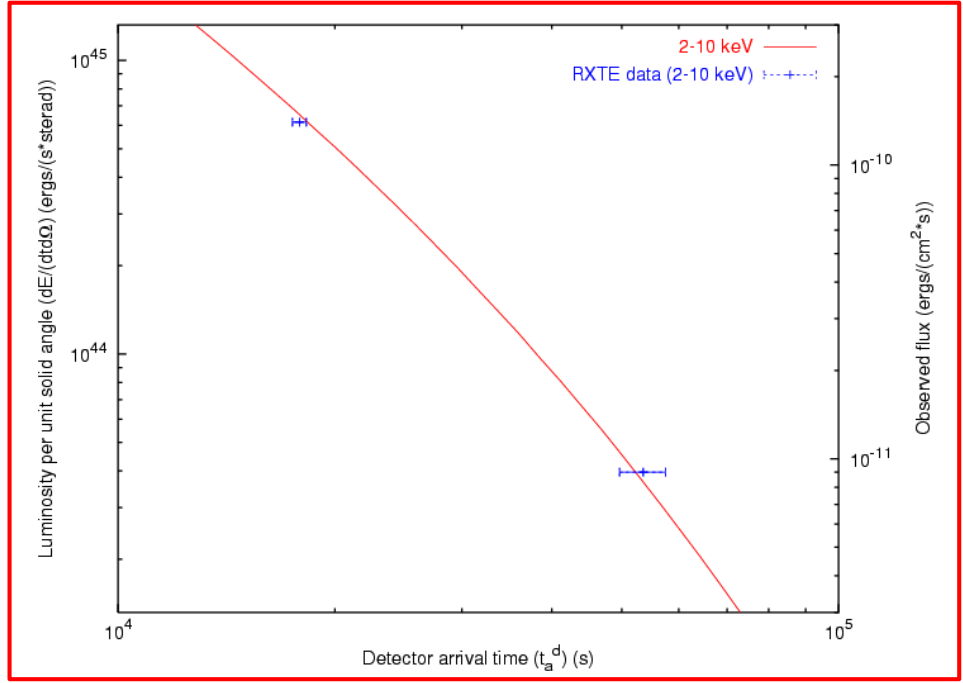} 
\caption{The perfect fit of the late part of the afterglow of our theoretical model for the $2$--$10$ keV energy bands. The data refers to the R-XTE observations (GCN Circ. 1996 \cite{gcn1996}). Details in Bernardini et al. \cite{b03mg10}.}
\label{030329af} 
\end{figure}

\begin{figure} 
\centering
\includegraphics[width=\hsize]{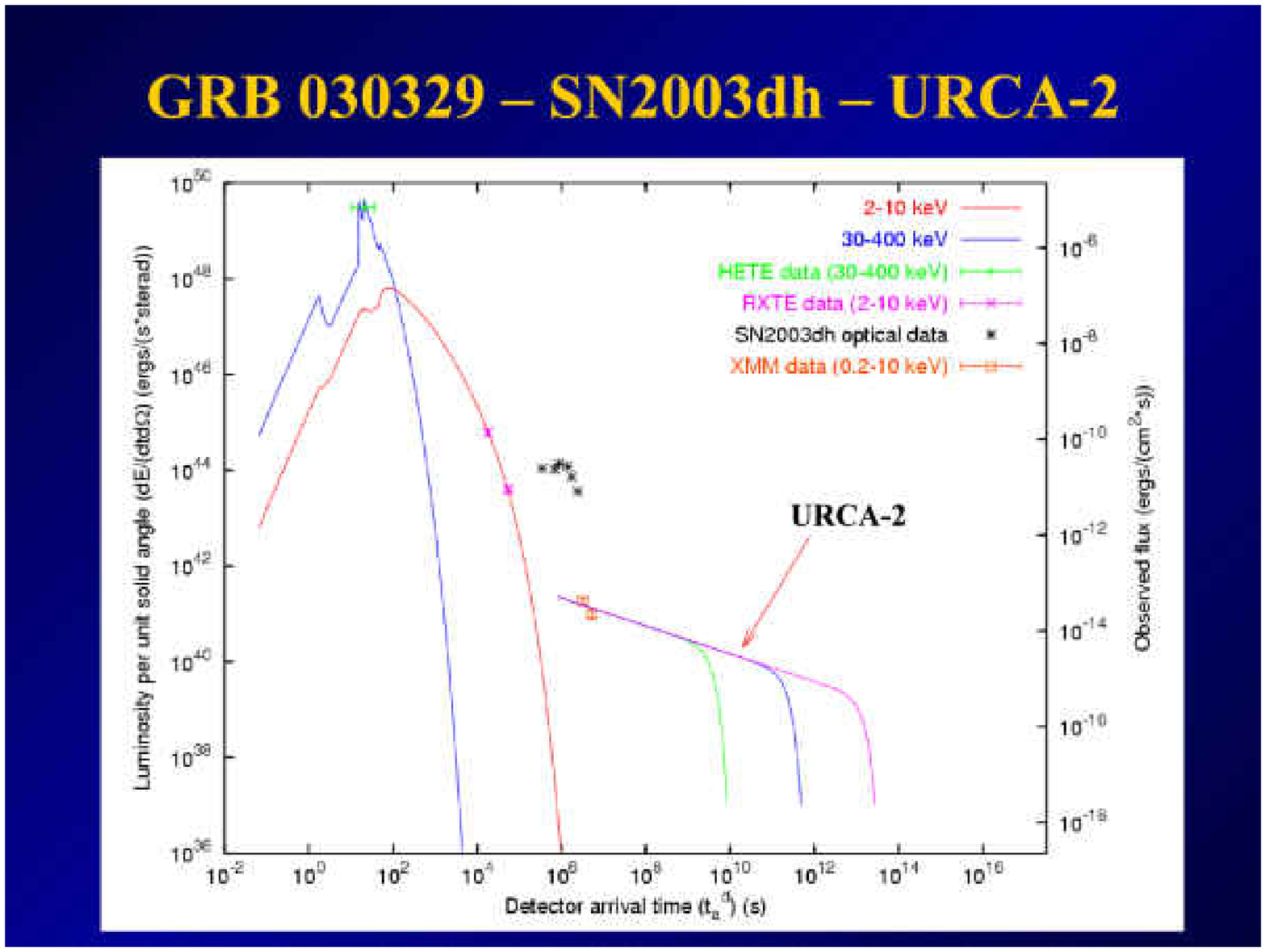} 
\caption{The dotted line represents our theoretically predicted GRB030329 light curve in $\gamma$-rays (30-400 keV) with the horizontal bar corresponding to the mean peak flux from HETE-2 (GCN Circ. 1997 \cite{gcn1997}). The solid line represents the corresponding one in X-rays (2-10 keV) with the experimental data obtained by R-XTE (GCN Circ. 1996 \cite{gcn1996}). The remaining points refer respectively to the optical VLT data (Hjorth et al. \cite{ha03}) of SN2003bw and to the X-ray XMM data (Tiengo et al. \cite{ta03}) of URCA-2. The dash-dotted lines corresponds to qualitative luminosity curves expected for URCA-2. It is interesting to compare and contrast these results with the ones for GRB980425/SN1998bw (see Fig. 3 in Ruffini et al. \cite{cospar02}). Details in Ruffini et al. \cite{030329}.}
\label{fig1} 
\end{figure}

We then obtain (see also Bernardini et al. \cite{b03mg10} in these proceedings) for the GRB 030329 the luminosities in given energy bands, computed in the range $2$--$400$ keV with very high accuracy. Figs. \ref{030329global}--\ref{fig1} shows the results for the luminosities in the $30$--$400$ keV and $2$--$10$ keV bands, including the ``prompt radiation''. Subsequently, the theoretically predicted GRB spectra have been evaluated at selected values of the arrival time (Ruffini et al. \cite{030329}). 

The splendid news received the evening before the presentation of this talk is graphically represented by the XMM observations shown in Fig. \ref{fig1}. Again, the XMM observations, like the corresponding ones of GRB 980425, occur after the decaying part of the afterglow and, in analogy to the one occurring in the system GRB 980425 / SN 1998bw / URCA-1, we call this source URCA-2. Further observations by XMM are highly recommended to follow the URCA-2 temporal evolution. Also in this system we are dealing with an astrophysical ``triptych'': GRB 030329 / SN 2003dh / URCA-2.

\section{Conclusions}

\subsection{On the GRB-Supernova connection}

We first stress some general considerations originating from comparing and contrasting the three GRB sources we have discussed:
\begin{enumerate}
\item The value of the $B$ parameter for all three sources occurs, as theoretically expected, in the allowed range (see Fig. \ref{bcross})
\begin{equation}
10^{-5} \le B \le 10^{-2}\, .
\label{brange}
\end{equation}
We have in fact:\\
\begin{center}
\begin{tabular}{ccc}
GRB 991216 & $B = 3.0\times 10^{-3}$ & $E_{dya} = 4.8 \times 10^{53}$ erg\\
GRB 980425 & $B = 7.0\times 10^{-3}$ & $E_{dya} = 1.1 \times 10^{48}$ erg\\
GRB 030329 & $B = 4.8\times 10^{-3}$ & $E_{dya} = 2.1 \times 10^{52}$ erg\\
\end{tabular}
\end{center}
\item The enormous difference in the GRB energy of the sources simply relates to the electromagnetic energy of the black hole given in Eq.(\ref{em}) which turns out to be smaller than the critical value given by Eq.(\ref{s1}). The fact that the theory is valid over $5$ orders of magnitude is indeed very satisfactory.
\item Also revealing is the fact that in both sources GRB 980425 and GRB 030329 the associated supernova energies are similar. We have, in fact, for both SN 1998bw and SN 2003dh an energy $\sim 10^{49}$ erg. Details in Fraschetti et al. \cite{f03mg10} and Bernardini et al. \cite{b03mg10}. The further comparison between the SN luminosity and the GRB intensity is crucial. In the case of GRB 980425 the GRB and the SN energies are comparable, and no dominance of one source over the other can be ascertained. In the case of GRB 030329 the energy of the GRB source is $10^3$ larger than the SN: in no way the GRB can originate from the SN event.
\end{enumerate}
The above stringent energetics considerations and the fact that GRBs occur also without an observed supernova give a strong evidence that GRBs cannot originate from supernovae.

\subsection{URCA-1 and URCA-2}

We turn now to the most exciting search for the nature of URCA-1 and URCA-2. We have already mentioned above that a variety of possibilities naturally appear. The first possibility is that the URCA sources are related to the black hole originating the GRB phenomenon. In order to probe such an hypothesis, it would be very important to find even a single case in which an URCA source occurs in association with a GRB and in absence of an associated supernova. Such a result, theoretically unexpected, would open an entire new problematic in relativistic astrophysics and in the physics of black holes.

If indeed, as we expect, the clear association between URCA sources and the supernovae occurring together with the GRBs, then it is clear that the analysis of the other two possibilities will be favored. Namely, an emission from processes occurring in the early phases of the expansion of the supernova remnant or the very exciting possibility that for the first time we are observing a newly born neutron star out of the supernova phenomenon. Of course, this last hypothesis is the most important one, since it would offer new fundamental information about the outcome of the gravitational collapse, about the equations of state at supranuclear densities and about a variety of fundamental issues of relativistic astrophysics of neutron stars. We shall focus in the following only on this last topic.

We have already recalled how the need for a rapid cooling process due to neutrino anti-neutrino emission in the process of gravitational collapse leading to the formation of a neutron star was considered for the first time by George Gamow and Mario Schoenberg in 1941 \cite{gs41}. It was Gamow who gave this process the name ``Urca process'', see \ref{gamow} and \ref{urca}. Since then, a systematic analysis of the theory of neutron star cooling was advanced by Tsuruta \cite{t64,t79}, Tsuruta and Cameron \cite{tc66}, Tsuruta et al. \cite{t02} and by Canuto \cite{c78}. The coming of age of X-ray observatories such as Einstein (1978-1981), EXOSAT (1983-1986), ROSAT (1990-1998), and the contemporary missions of Chandra and XMM-Newton since 1999 dramatically presented an observational situation establishing very embarrassing and stringent upper limits to the surface temperature of neutron stars in well known historical supernova remnants (see e.g. Romani \cite{r87}). It was so that, for some remnants, notably SN 1006 and the Tycho supernova, the upper limits to the surface temperatures were significantly lower than the temperatures given by standard cooling times (see e.g. Romani \cite{r87}). Much of the theoretical works has been mainly directed, therefore, to find theoretical arguments in order to explain such low surface temperature $T_s \sim 0.5$--$1.0\times 10^6$ K --- embarrassingly low, when compared to the initial hot ($\sim 10^{11}$ K) birth of a neutron star in a supernova explosion (see e.g. Romani \cite{r87}). Some very important steps in this direction of research have been represented by the works of Van Riper \cite{vr88,vr91}, Lattimer and his group \cite{bl86,lvrpp94} and by the most extensive work of Yakovlev and his group \cite{yp04}. The youngest neutron star to be searched for using its thermal emission in this context has been the pulsar PSR J0205+6449 in 3C 58 (see e.g. Yakovlev and Pethick \cite{yp04}), which is $820$ years old! Recently, evidence for the detection of thermal emission from the crab nebula pulsar was reported by Trumper \cite{t05} which is, again, $951$ years old.

In the case of URCA-1 and URCA-2, we are exploring a totally different regime: the X-ray emission possibly from a recently born neutron star in the first days -- months of its existence, where no observations have yet been performed and no embarrassing constraints upper limits on the surface temperature exist. The reason of approaching first the issue of the thermal emission from the neutron star surface is extremely important, since in principle it can give information on the equations of state in the core at supranuclear densities and on the detailed mechanism of the formation of the neutron star itself and the related neutrino emission. It is of course possible that the neutron star is initially fast rotating and its early emission is dominated by the magnetospheric emission or by accretion processes from the remnant which would overshadow the thermal emission. In that case a periodic signal related to the neutron star rotational period should in principle be observable in a close enough GRB source provided the suitable instrumentation from the Earth.

The literature on young born neutron star is relatively scarse today. There are some very interesting contributions which state: ``The time for a neutron star's center to cool by the direct URCA process to a temperature $T$ has been estimated to be $t = 20 \left[T/\left(10^9 K\right)\right]^{-4}$ s. The direct URCA process and all the exotic cooling mechanisms only occur at supranuclear densities. Matter at subnuclear densities in neutron star crust cools primarily by diffusion of heat to the interior. Thus the surface temperature remains high, in the vicinity of $10^6$ K or more, until the crust's heat reservoir is consumed. After this diffusion time, which is on the order of $1$--$100$ years, the surface temperature abruptly plunges to values below $5 \times 10^5$ K'' (Lattimer et al. \cite{lvrpp94}). ``Soon after a supernova explosion, the young neutron star has large temperature gradients in the inner part of the crust. While the powerful neutrino emission quickly cools the core, the crust stays hot. The heat gradually flows inward on a conduction time scale and the whole process can be thought of as a cooling wave propagation from the center toward the surface'' (Gnedin et al. \cite{ya01}).

The two considerations we have quoted above are developed in the case of spherical symmetry and we would like to keep the mind open, in this new astrophysical field, to additional factors, some more traditional than others, to be taken into account. Among the traditional ones we recall: 1) the presence of rotation and magnetic field which may affect the thermal conductivity and the structure of the surface, as well as the above mentioned magnetospheric emission; 2) there could be accretion of matter from the expanding nebula; and, among the nontraditional ones, we recall 3) some exciting theoretical possibilities advanced by Dyson on volcanoes on neutron stars \cite{d69} as well as iron helide on neutron star \cite{d71}, as well as the possibility of piconuclear reactions on neutron star surface discussed in Lai \& Salpeter \cite{ls97}.

All the above are just scientific arguments to attract attention on the abrupt fall in luminosity reported in this meeting on URCA-1 by Elena Pian which is therefore, in this light, of the greatest scientific interest and further analysis should be followed to check if a similar behavior will be found in future XMM and Chandra observations also in URCA-2.

\subsection{Astrophysical implications}

In addition to these very rich problematics in the field of theoretical physics and theoretical astrophysics, there are also more classical astronomical and astrophysical issues, which will need to be answered if indeed the observations of a young neutron star will be confirmed. An important issue to be addressed will be how the young neutron star can be observed, escaping from being buried under the expelled matter of the collapsing star. A possible explanation can originate from the binary nature of the newly born neutron star: the binary system being formed by the newly formed black hole and the triggered gravitational collapse of a companion evolved star leading, possibly, to a ``kick'' on and ejection of the newly born neutron star. Another possibility, also related to the binary nature of the system, is that the supernova progenitor star has been depleted of its outer layer by dynamic tidal effects.

In addition, there are other topics in which our scenario can open new research directions in fundamental physics and astrophysics:\\
1) The problem of the instability leading to the complete gravitational collapse of a $\sim 10M_\odot$ star needs the introduction of a new critical mass for gravitational collapse, which is quite different from the one for white dwarfs and neutron stars which has been widely discussed in the current literature (see e.g. Giacconi \& Ruffini \cite{gr78}).\\
2) The issue of the trigger of the instability of gravitational collapse induced by the GRB on the progenitor star of the supernova or, vice versa, by the supernova on the progenitor star of the GRB needs accurate timing and the considerations of new relativistic phenomena.\\
3) The general relativistic instability induced on a nearby star by the formation of a black hole needs some very basic new developments in the field of general relativity.

Only a very preliminary work exists on this subject, by Jim Wilson and his collaborators, see e.g. the paper by Mathews and Wilson in these proceedings \cite{mw}. The reason for the complexity in answering such a question is simply stated: unlike the majority of theoretical work on black holes, which deals mainly with one-body solutions, we have to address here a many-body problem in general relativity. We are starting in these days to reconsider, in this framework, some classic work by Fermi \cite{f21}, Hanni and Ruffini \cite{hr73}, Majumdar \cite{m47}, Papapetrou \cite{p47}, Parker et al. \cite{p73}, Bini et al. \cite{bgr04} which may lead to a new understanding of general relativistic effects relevant to these astrophysical ``triptychs''.

\section*{Acknowledgments}
We are thankful to Rashid Sunyaev, Lev Titarchuk, Jim Wilson and Dima Yakovlev for many interesting theoretical discussions, as well as to Lorenzo Amati, Lucio Angelo Antonelli, Enrico Costa, Filippo Frontera, Luciano Nicastro, Elena Pian, Luigi Piro, Marco Tavani and all the BeppoSAX team for assistance in the data analysis.

\appendix

\section{On the Urca process}\label{gamow}

From G. Gamow \cite{g70}:

``The summer of 1939 I spent with my family vacation on the Copacabana beach in Rio de Janeiro. One evening, visiting the famous Casino da Urca to watch the gamblers, I was introduced to a young theoretical physicist born on an Amazon River plantation, named Mario Schoenberg. We became friends, and I arranged for him a Guggenheim fellowship to spend a year in Washington to work with me in nuclear astrophysics. His visit was very successful, and we hit upon a process which could be responsible for the vast stellar explosions known as supernovae. The trick is done by alternative absorption and reemission of one of the thermal electrons in the very hot (billions of degrees!) stellar interior by various atomic nucleai. Both processes are accompanied by the emission of neutrinos and antineutrinos which, possessing tremendous penetrating power, pass through the body of a star like a swarm of musquitoes through chicken wire and carry with them large amount of energy. Thus, the stellar interior cools rapidly, the pressure drops, and the stellar body collapse with a great explosion of light and heat.

All this is too complicated to explain in nontechnical words, and I am mentioning it only as background for how we came to give that process its name. We called it the Urca process, partially to commemorate the casino in which we first met, and partially because the Urca process results in a rapid disappearance of thermal energy from the interior of the star, similar to the rapid disappearance of money from the pockets of the gamblers of the Casino da Urca. Sending our article ``On the Urca process'' for publication in {\em The Physical Review} I was worried that the Editor would ask why we called the process ``Urca''. After much thought I decided to say that this is short for ``UnRecordable Cooling Agent'', but they never asked. Today, there are other known cooling processes involving neutrinos which work even faster than the Urca process. For example, a neutrino pair can be formed instead of two gamma quanta in the annihilation of a positive and negative electrons''.

\section{Casino da Urca today}\label{urca}

\includegraphics[width=\hsize,clip]{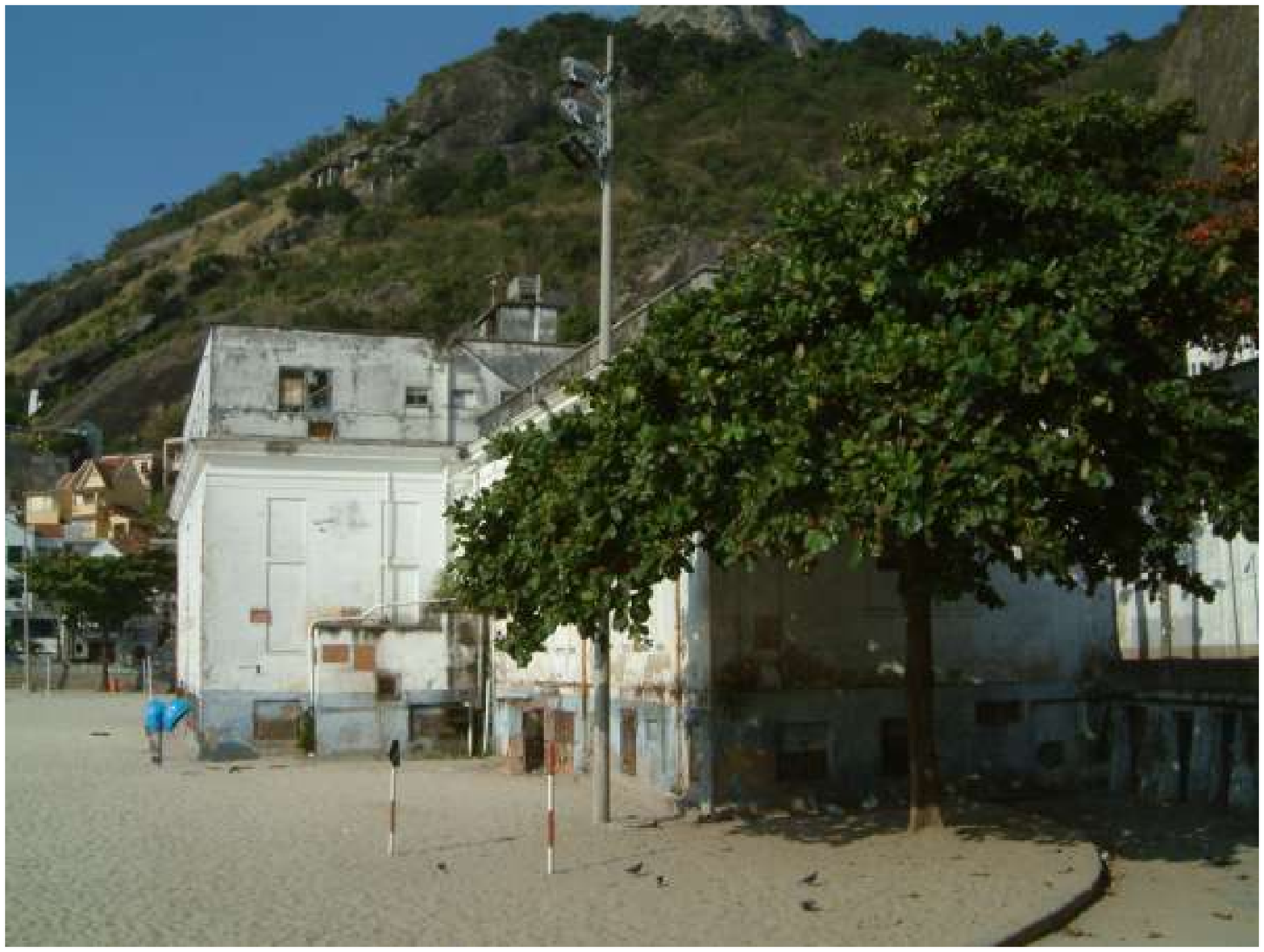}


\begin{thebibliography}{99}

\bibitem{msch}
M. Schwarzschild, ``Structure and evolution of the stars'', Dover Publications (New York, 1965).

\bibitem{hbpsc68}
A. Hewish, S.J. Bell, J.D. Pilkington, P.F. Scott, R.A. Collins, {\it Nature}, {\bf 217}, 709 (1968).

\bibitem{gam36}
G. Gamow, ``Atomic nuclei and nuclear transformations'', second edition, Oxford (1936).

\bibitem{g68}
T. Gold, {\it Nature}, {\bf 218}, 731 (1968).

\bibitem{g69}
T. Gold, {\it Nature}, {\bf 221}, 27 (1969).

\bibitem{gr78}
R. Giacconi, R. Ruffini (eds.), ``Physics and astrophysics of neutron stars and black holes'', North-Holland (Amsterdam, 1978).

\bibitem{rw71}
R. Ruffini, J.A. Wheeler, {\it Phys. Today}, {\bf 24}, 30 (1971).

\bibitem{lr73}
R. Leach, R. Ruffini, {\it ApJ}, {\bf 180}, L15 (1973).

\bibitem{ll2}
L.D. Landau, E.M. Lifshitz, ``The classical theory of fields'', fourth revised edition, Butterworth-Heinemann (Oxford, 2003), p. 352.

\bibitem{rr74}
C. Rhoades, R. Ruffini, {\it Phys. Rev. Lett.}, {\bf 32}, 324 (1974).

\bibitem{cr71}
D. Christodoulou, R. Ruffini, {\it Phys. Rev. D}, {\bf 4}, 3552 (1971).

\bibitem{lett1} 
R. Ruffini, C.L. Bianco, P. Chardonnet, F. Fraschetti, S.-S. Xue, {\it ApJ}, {\bf 555}, L107 (2001).

\bibitem{lett2} 
R. Ruffini, C.L. Bianco, P. Chardonnet, F. Fraschetti, S.-S. Xue, {\it ApJ}, {\bf 555}, L113 (2001).

\bibitem{lett3} 
R. Ruffini, C.L. Bianco, P. Chardonnet, F. Fraschetti, S.-S. Xue, {\it ApJ}, {\bf 555}, L117 (2001).

\bibitem{cospar02} 
R. Ruffini, M.G. Bernardini, C.L. Bianco, P. Chardonnet, F. Fraschetti, S.-S. Xue, {\it Adv. Sp. Res.}, {\bf 34}, 2715 (2004).

\bibitem{f03mg10}
F. Fraschetti, M.G. Bernardini, C.L. Bianco, P. Chardonnet, R. Ruffini, S.-S.Xue, in these proceedings.

\bibitem{s75}
I.B. Strong, in ``Neutron Stars, Black Holes and Binary X-Ray Sources'', H. Gursky and R. Ruffini (eds.), D. Reidel Publishing Company (Utrecht, 1975).

\bibitem{batse4b}
W.S. Paciesas, et al., {\it ApJ Suppl.}, {\bf 122}, 465 (1999).

\bibitem{ka93}
C. Kouveliotou, et al., {\it ApJ}, {\bf 413}, L101 (1993).

\bibitem{t98}
M. Tavani, {\it ApJ}, {\bf 497}, L21 (1998).

\bibitem{ca97}
E. Costa, et al., {\it Nature}, {\bf 387}, 783 (1997).

\bibitem{dr75}
T. Damour, R. Ruffini, {\it Phys. Rev. Lett.}, {\bf 35}, 463 (1975).

\bibitem{e05}
A. Einstein, {\it Ann. Phys. (Germany)}, {\bf 17}, 891 (1905)

\bibitem{he35}
W. Heisenberg, H. Euler, {\it Zeits. Phys.}, {\bf 98}, 714 (1935).

\bibitem{s51}
J. Schwinger, {\it Phys. Rev.}, {\bf 82}, 664 (1951).

\bibitem{rvx05}
R. Ruffini, L. Vitagliano, S.-S. Xue, {\it Phys. Rep.}, in preparation.

\bibitem{Brasile} 
R. Ruffini, C.L. Bianco, P. Chardonnet, F. Fraschetti, L. Vitagliano, S.-S. Xue, ``New Perspectives in Physics and Astrophysics from the Theoretical Understanding of Gamma-Ray Bursts'', in {\it COSMOLOGY AND GRAVITATION: X$^{th}$ Brazilian School of Cosmology and Gravitation; 25$^{th}$ Anniversary (1977-2002)}, M. Novello, S.E. Perez-Bergliaffa (eds.), {\it AIP Conf. Proc.}, {\bf 668}, 16 (2003).

\bibitem{p99}
T. Piran, {\it Phys. Rep.}, {\bf 314}, 575 (1999).

\bibitem{p00}
T. Piran, {\it Phys. Rep.}, {\bf 333}, 529 (2000).

\bibitem{rm94}
M.J. Rees, P. M\'esz\'aros, {\it ApJ}, {\bf 430}, L93 (1994).

\bibitem{px94}
B. Paczy\'nski, G. Xu, {\it ApJ}, {\bf 427}, 708 (1994).

\bibitem{sp97}
R. Sari, T. Piran, {\it ApJ}, {\bf 485}, 270 (1997).

\bibitem{f99}
E.E. Fenimore, {\it ApJ}, {\bf 518}, 375 (1999).

\bibitem{fcrsyn99}
E.E. Fenimore, C. Cooper, E. Ramirez-Ruiz, M.C. Sumner, A. Yoshida, M. Namiki, {\it ApJ}, {\bf 512}, 683 (1999).

\bibitem{r70}
R. Ruffini, ``On the energetics of Black Holes'', in ``Black Holes - Les astres occlus'', C. and B.S. De Witt (eds.), Gordon and Breach (New York, 1973).

\bibitem{kerr}
R.P. Kerr, {\it Phys. Rev. Lett.}, {\bf 11}, 237 (1963).

\bibitem{newman}
E.T. Newman, E. Couch, K. Chinnapared, A. Exton, A. Prakash, R. Torrence, {\it J. Mat. Phys.}, {\bf 6}, 918 (1965).

\bibitem{carter}
B. Carter, {\it Phys. Rev.}, {\bf 174}, 1559 (1968).

\bibitem{rrw}
M. Rees, R. Ruffini, J.A. Wheeler, ``Black holes, gravitational waves and cosmology'', Gordon and Breach (New York, 1974).

\bibitem{ckf}
B. Carter, in ``Kerr fest'', Cambridge University Press, in press.

\bibitem{bcjr}
D. Bini, C. Cherubini, R.T. Jantzen, R. Ruffini, {\it Prog. Theo. Phys.}, {\bf 107}, 967 (2002).

\bibitem{b33}
M. Born, {\it Proc. Roy. Soc. London A}, {\bf 143}, 410 (1933).

\bibitem{cr78} 
G. Cavallo, M.J. Rees, {\it MNRAS}, {\bf 183}, 359 (1978).

\bibitem{ch81}
G. Cavallo, H.M. Horstman, {\it Astroph. Sp. Sci.}, {\bf 75}, 117 (1981).

\bibitem{hc83}
H.M. Horstman, G. Cavallo, {\it A\&A}, {\bf 122}, 119 (1983).

\bibitem{rukyoto}
R. Ruffini, ``Beyond the Critical Mass: The Dyadosphere of Black Holes'', in {\it Black Holes and High Energy Astrophysics, Proceedings of the $49^{th}$ Yamada Conference}, H. Sato and N. Sugiyama (eds.), Universal Ac. Press (Tokyo, 1998). 

\bibitem{prx98} 
G. Preparata, R. Ruffini, S.-S. Xue, {\it A\&A}, {\bf 338}, L87 (1998).

\bibitem{rvx03a}
R. Ruffini, L. Vitagliano, S.-S. Xue, {\it Phys. Lett. B}, {\bf 559}, 12 (2003).

\bibitem{crv02}
C. Cherubini, R. Ruffini, L. Vitagliano, {\it Phys. Lett. B}, {\bf 545}, 226 (2002).

\bibitem{rv02a}
R. Ruffini, L. Vitagliano, {\it Phys. Lett. B}, {\bf 545}, 233 (2002).

\bibitem{rv02b}
R. Ruffini, L. Vitagliano, {\it IJMPD}, {\bf 12}, 121 (2003).

\bibitem{rvx03b}
R. Ruffini, L. Vitagliano, S.-S. Xue, {\it Phys. Lett. B}, {\bf 573}, 33 (2003).

\bibitem{rfvx05} 
R. Ruffini, F. Fraschetti, L. Vitagliano, S.-S. Xue, {\it IJMPD}, in press.

\bibitem{rswx99}
R. Ruffini, J.D. Salmonson, J.R. Wilson, S.-S. Xue, {\it A\&A}, {\bf 350}, 334 (1999).

\bibitem{psn93} 
T. Piran, A. Shemi, R. Narayan, {\it MNRAS}, {\bf 263}, 861 (1993).

\bibitem{bm95} 
G.S. Bisnovatyi-Kogan, M.V.A. Murzina, {\it Phys. Rev. D}, {\bf 52}, 4380 (1995).

\bibitem{mlr93} 
P. M\'esz\'aros, P. Laguna, M.J. Rees, {\it ApJ}, {\bf 415}, 181 (1993).

\bibitem{rswx00}
R. Ruffini, J.D. Salmonson, J.R. Wilson, S.-S. Xue, {\it A\&A}, {\bf 359}, 855 (2000).

\bibitem{sp90}
A. Shemi, T. Piran, {\it ApJ}, {\bf 365}, L55 (1990).

\bibitem{bfrvx}
C.L. Bianco, F. Fraschetti, R. Ruffini, S. Vincent, S.-S. Xue, in preparation.

\bibitem{cd99}
J. Chiang, C.D. Dermer, {\it ApJ}, {\bf 512}, 699 (1999).

\bibitem{bm76}
R.D. Blandford, C.F. McKee, {\it Phys. Fluids}, {\bf 19}, 1130 (1976).

\bibitem{s97}
R. Sari, {\it ApJ}, {\bf 489}, L37 (1997).

\bibitem{s98}
R. Sari, {\it ApJ}, {\bf 494}, L49 (1998).

\bibitem{w97}
E. Waxman, {\it ApJ}, {\bf 491}, L19 (1997).

\bibitem{rm98}
M.J. Rees, P. M\'esz\'aros, {\it ApJ}, {\bf 496}, L1 (1998).

\bibitem{gps99}
J. Granot, T. Piran, R. Sari, {\it ApJ}, {\bf 513}, 679 (1999).

\bibitem{pm98c}
A. Panaitescu, P. M\'esz\'aros, {\it ApJ}, {\bf 493}, L31 (1998).

\bibitem{gw99}
A. Gruzinov, E. Waxman, {\it ApJ}, {\bf 511}, 852 (1999).

\bibitem{vpkw00}
J. van Paradijs, C. Kouveliotou, R.A.M.J. Wijers, {\it Annu. Rev. Astron. \& Astroph.}, {\bf 38}, 379 (2000).

\bibitem{m02}
P. M\'esz\'aros, {\it Annu. Rev. Astron. \& Astroph.}, {\bf 40}, 137 (2002).

\bibitem{c39}
P. Couderc, {\it Ann. Astroph.}, {\bf 2}, 271 (1939).

\bibitem{EQTS_ApJL}
C.L. Bianco, R. Ruffini, {\it ApJ}, {\bf 605}, L1 (2004).

\bibitem{EQTS_ApJL2}
C.L. Bianco, R. Ruffini, {\it ApJ}, {\bf 620}, L23 (2005).

\bibitem{rbcfx02_letter}
R. Ruffini, C.L. Bianco, P. Chardonnet, F. Fraschetti, S.-S. Xue, {\it ApJ}, {\bf 581}, L19 (2002).

\bibitem{pls}
R. Ruffini, P.Chardonnet, C.L. Bianco, S.-S. Xue, F. Fraschetti, in {\it Pour la science}, n. 294 (April 2002).

\bibitem{Spectr2}
R. Ruffini, C.L. Bianco, P. Chardonnet, F. Fraschetti, V. Gurzadyan, S.-S. Xue, {\it IJMPD}, in press.

\bibitem{ll6}
L.D. Landau, E.M. Lifshitz, ``Fluid Mechanics'', Butterworth-Heinemann (Oxford, 2003).

\bibitem{zr66}
Ya.B. Zel'dovich, Yu.P. Rayzer, ``Physics of shock waves and high-temperature hydrodynamic phenomena'', Wallace D. Hayes and Ronald F. Probstein (eds.), Academic press (New York and London, 1966).

\bibitem{sedov}
L.I. Sedov, ``Similarity and Dimensional Methods in Mechanics'', Academic Press (New York, 1959).

\bibitem{mc75}
C.M. McKee, L.L Cowie, {\it ApJ}, {\bf 195}, 715 (1975).

\bibitem{tt91}
G. Tenorio-Tagle, M. Rozyczka, J. Franco, P. Bodenheimer, {\it MNRAS}, {\bf 251}, 318 (1991).

\bibitem{sn92}
J.M. Stone, M.L. Norman, {\it ApJ Suppl.}, {\bf 80}, 753 (1992).

\bibitem{j99}
B. Jun, T.W. Jones, {\it ApJ}, {\bf 511}, 774 (1999).

\bibitem{gcg02}
G. Ghirlanda, A. Celotti, G. Ghisellini, {\it A\&A}, {\bf 393}, 409 (2002).

\bibitem{pa98}
R.D. Preece, M.S. Briggs, R.S. Mallozzi, G.N. Pendleton, W.S. Paciesas, D.L. Band, {\it ApJ}, {\bf 506}, L23 (1998)

\bibitem{Spectr1}
R. Ruffini, C.L. Bianco, P. Chardonnet, F. Fraschetti, V. Gurzadyan, S.-S. Xue, {\it IJMPD}, {\bf 13}, 843 (2004).

\bibitem{fa00}
F. Frontera, et al, {\it ApJ Suppl.}, {\bf 127}, 59 (2000).

\bibitem{p99b}
L. Piro, et al., {\it ApJ}, {\bf 514}, L73 (1999).

\bibitem{b93}
D. Band, et al., {\it ApJ}, {\bf 413}, 281 (1993).

\bibitem{p98}
B. Paczy\'nski, {\it ApJ}, {\bf 494}, L45 (1998).

\bibitem{ka98}
S.R. Kulkarni, et al., {\it Nature}, {\bf 395}, 663 (1998).

\bibitem{ia98}
K. Iwamoto, et al., {\it Nature}, {\bf 395}, 672 (1998).

\bibitem{pian00} 
E. Pian, et al. {\it ApJ}, {\bf 536}, 778 (2000).

\bibitem{i99}
K. Iwamoto, {\it ApJ}, {\bf 512}, L47 (1999).

\bibitem{b03mg10}
M.G. Bernardini, C.L. Bianco, P. Chardonnet, F. Fraschetti, R. Ruffini, S.-S.Xue, in these proceedings.

\bibitem{030329}
R. Ruffini, M.G. Bernardini, C.L. Bianco, P. Chardonnet, F. Fraschetti, S.-S. Xue, in preparation.

\bibitem{gcn1996} 
GCN Circ. 1996 (2003).

\bibitem{gcn1997} 
GCN Circ. 1997 (2003).

\bibitem{ha03} 
J. Hjorth, et al. {\it Nature}, {\bf 423}, 847 (2003).

\bibitem{ta03}
A. Tiengo, S. Mereghetti, G. Ghisellini, E. Rossi, G. Ghirlanda, N. Schartel, {\it A\&A}, {\bf 409}, 983 (2003).

\bibitem{gs41}
G. Gamow, M. Schoenberg, {\it Phys. Rev.}, {\bf 59}, 539 (1941).

\bibitem{t64}
S. Tsuruta, {\it Ph.D. thesis}, Columbia University (1964).

\bibitem{t79}
S. Tsuruta, {\it Phys. Rep.}, {\bf 56}, 237 (1979).

\bibitem{tc66}
S. Tsuruta, A.G.W. Cameron, {\it Can. J. Phys.}, {\bf 44}, 1863 (1966).

\bibitem{t02}
S. Tsuruta, M.A. Teter, T. Takatsuka, T. Tatsumi, R. Tamagaki, {\it ApJ}, {\bf 571}, L143 (2002).

\bibitem{c78}
V. Canuto, in ``Physics and astrophysics of neutron stars and black holes'', R. Giacconi, R. Ruffini (eds.), North-Holland (Amsterdam, 1978).

\bibitem{r87}
R.W. Romani, {\it ApJ}, {\bf 313}, 718 (1987).

\bibitem{vr88}
K.A. Van Riper, {\it ApJ}, {\bf 329}, 339 (1988).

\bibitem{vr91}
K.A. Van Riper, {\it ApJ Suppl.}, {\bf 75}, 449 (1991).

\bibitem{bl86}
A. Burrows, J.M. Lattimer, {\it ApJ}, {\bf 307}, 178 (1986).

\bibitem{lvrpp94}
J.M. Lattimer, K.A. Van Riper, M. Prakash, M. Prakash, {\it ApJ}, {\bf 425}, 802 (1994).

\bibitem{yp04}
D.G. Yakovlev, C.J. Pethick, {\it Ann. Rev. Astron. Astroph.}, {\bf 42}, 169 (2004).

\bibitem{t05}
J.E. Trumper, in {\it ASI proceedings of The Electromagnetic Spectrum of Neutron Stars}, in press, \verb#astro-ph/0502457#

\bibitem{ya01}
O.Y. Gnedin, D.G. Yakovlev, A.Y. Potekhin, {\it MNRAS}, {\bf 324}, 725 (2001).

\bibitem{d69}
F. Dyson, {\it Nature}, {\bf 223}, 486 (1969).

\bibitem{d71}
F. Dyson, {\it Ann. Phys.}, {\bf 63}, 1 (1971).

\bibitem{ls97}
D. Lai, E.E. Salpeter, {\it ApJ}, {\bf 491}, 270 (1997).

\bibitem{mw}
R. Mathews, J. Wilson, in these proceedings.

\bibitem{f21}
E. Fermi, {\it Il Nuovo Cimento}, {\bf 22}, 176 (1921).

\bibitem{hr73}
R. Hanni, R. Ruffini, {\it Phys. Rev. D}, {\bf 8}, 3259 (1973).

\bibitem{m47}
S.D. Majumdar, {\it Phys. Rev.}, {\bf 72}, 390 (1947).

\bibitem{p47}
A. Papapetrou, {\it Proc. R. Irish Acad.}, {\bf 51}, 191 (1947).

\bibitem{p73}
L. Parker, R. Ruffini, D. Wilkins, {\it Phys. Rev. D}, {\bf 7}, 2874 (1973).

\bibitem{bgr04}
D. Bini, A. Geralico, R. Ruffini, in preparation.

\bibitem{g70}
G. Gamow, ``My wordlines - an informal autobiography'', Viking press (New York, 1970).

\end{thebibliography}
\end{document}